\documentclass[prd,showpacs,superscriptaddress,nofootinbib,floatfix,preprint,11pt]{revtex4}
 
\usepackage{graphicx}
\usepackage{amsmath}
\usepackage{amsfonts}    
\usepackage{mathrsfs}
\usepackage{array} 
\usepackage[%
    pdftitle={Black holes in the conical ensemble},%
    pdfauthor={Daniel Grumiller, Robert McNees, Simone Zonetti},%
    pdfkeywords={Black hole thermodynamics, conical singularities, Euclidean path integral, two-dimensional dilaton gravity, string theory in two dimensions, spherically symmetric black holes, BTZ black hole},
]{hyperref}
  
\newcommand{\eq}[2]{\begin{equation} #1 \label{#2} \end{equation}}

\newcommand{\be}{\beta}

\newcommand{\de}{\delta}

\newcommand{\la}{\lambda}

\newcommand{\Ga}{\Gamma}
\newcommand{\De}{\Delta}

\newcommand\MM{\mathcal{M}}
\newcommand\dM{\partial \MM}
\newcommand\ZZ{\mathcal{Z}}
\newcommand\CurlD{\mathscr{D}}
\newcommand\nts{\negthickspace}
\newcommand\ns{\negthickspace}
\newcommand\bns{\nts \nts \nts}
\newcommand\defeq{\mathrel{\mathop:}=}
\newcommand{\dd}{\mathrm{d}} 
\newcommand{\deficit}{\alpha}
\newcommand{\ts}[1]{\textrm{\tiny #1}}
\newcommand{\ms}[1]{\textrm{\tiny $#1$}}

\begin{document}

\title{Black holes in the conical ensemble \\ 
\normalsize{(\emph{or}, Getting deficits under control)}}

\author{Daniel Grumiller} \email{grumil@hep.itp.tuwien.ac.at}
\affiliation{Institute for Theoretical Physics, Vienna University of Technology, Wiedner Hauptstrasse 8-10/136, A-1040 Vienna, Austria}

\author{Robert McNees} \email{rmcnees@luc.edu}
\affiliation{Loyola University Chicago, Department of Physics, Chicago, IL 60660}

\author{Simone Zonetti} \email{simone.zonetti@uclouvain.be}
\affiliation{Centre for Cosmology, Particle Physics and Phenomenology (CP3), Institut de Recherche en Math\'ematique et Physique, Universit\'e catholique de Louvain, Chemin du Cyclotron 2, B-1348, Louvain-la-Neuve, Belgium}

\date{\today}

\preprint{TUW--12--30, {\tt arXiv:~1210.xxxx}}

\begin{abstract}
We consider black holes in an ``unsuitable box'': a finite cavity coupled to a thermal reservoir at a temperature different than the black hole's Hawking temperature. These black holes are described by metrics that are continuous but not differentiable due to a conical singularity at the horizon. We include them in the Euclidean path integral sum over configurations, and analyze the effect this has on black hole thermodynamics in the canonical ensemble. Black holes with a small deficit (or surplus) angle may have a smaller internal energy or larger density of states than the nearby smooth black hole, but they always have a larger free energy. Furthermore, we find that the ground state of the ensemble never possesses a conical singularity. When the ground state is a black hole, the contributions to the canonical partition function from configurations with a conical singularity are comparable to the contributions from smooth fluctuations of the fields around the black hole background. Our focus is on highly symmetric black holes that can be treated as solutions of two-dimensional dilaton gravity models: examples include Schwarzschild, asymptotically Anti-de Sitter, and stringy black holes. 
\end{abstract}


\maketitle

\tableofcontents

\section{Introduction}
\label{sec:Intro}

Black hole thermodynamics has a variety of applications, from the insights of quantum gravity Gedankenexperiments to practical calculations in heavy ion collisions and condensed matter physics. And there are just as many formalisms for describing the thermodynamic behavior: the classical four laws of black hole mechanics, quantum fields on curved backgrounds, microstate counting by virtue of the Cardy formula, and the Euclidean path integral formulation, to name a few. Each approach has its benefits, but the path integral approach has proven especially useful for practical calculations at 0- and 1-loop level, particularly in the context of gauge/gravity dualities. This is due, in part, to its direct connection with the classical formulation of the gravitational theory. Given an action $I_{E}$ for the theory, every relevant field configuration contributes with weight $\exp(-I_{E})$. Of course, the identification and enumeration of the `relevant field configurations' may present a challenge.

Already in the early days of path integrals it was discovered that the most relevant class of paths behave like the `Weierstrass monster': they are continuous but not differentiable at any point. In a simple system, like a point particle in certain potentials, the contributions from these paths can be accounted for. But the prospect of summing contributions from non-differentiable field configurations seems overwhelming for a theory as complex as gravity. Instead, calculations that employ the gravitational path integral will often focus exclusively on the contributions from geometries that satisfy some smoothness conditions. This assumption may be justified by physically reasonable results, but the fact remains that smooth metrics account for only a fraction of the support of the path integral measure. The alternative is to require only continuity of the metric when performing path integral calculations, while relaxing the requirement of differentiability.

In this paper we make some modest progress in this direction by including metrics with conical singularities in the Euclidean path integral for a number of gravitational theories. Specifically, we consider the Euclidean partition function for a canonical ensemble defined inside a finite cavity. Field configurations in the ensemble satisfy certain boundary conditions at the wall of the cavity, where the system is coupled to an external thermal reservoir. In some cases (theories with asymptotically anti-de Sitter boundary conditions, for instance) the walls of the box can be removed to an asymptotic region without compromising the existence of the ensemble, but this is not always possible. Usually one computes the partition function for this ensemble by summing contributions from metrics that are regular everywhere. We will relax this condition and include certain configurations with a conical singularity. This ensemble could be referred to as the \emph{conical ensemble}, to distinguish it from the usual canonical ensemble. We then pose the following questions:
\begin{quote}
	How do configurations with conical singularities contribute to the partition function? Can the ground state of the ensemble have a conical singularity?
\end{quote}
These questions can be answered quite generally for two-dimensional dilaton gravity. This class of models includes the spherically symmetric reduction of higher-dimensional theories that admit Schwarzschild, Schwarzschild-AdS, Reissner-Nordstr\"om, and BTZ black holes as solutions, as well as target space actions associated with certain string theory black holes.

How are conical singularities incorporated into the ensemble? In the semiclassical approximation the partition function is dominated by solutions of the classical equations of motion, with sub-leading contributions coming from smooth fluctuations of the fields around these configurations. We now wish to take into account geometries that are regular everywhere except for a single point. The dominant contributions would seem to come from configurations that ``almost'' extremize the action: they satisfy the equations of motion everywhere except for the singular point, similar to the `off-shell black holes' considered by Carlip and Teitelboim \cite{Carlip:1993sa}. The assumptions of our framework then require this point to sit either at the center of the cavity, or the horizon of a black hole. From the point of view of a higher dimensional model, a conical singularity at any other point other than the center of the cavity would not be consistent with the spherically symmetric reduction of the action. In the context of the two-dimensional dilaton gravity, the fact that the configurations satisfy the equations of motion everywhere except a single point implies the existence of a certain Killing vector that forces the singularity to sit at the center of the cavity. Thus, the dominant contributions to the partition function from configurations with a conical singularity correspond to solutions of the equations of motion that break down at the horizon, where there is a delta-function singularity in the curvature.

In the canonical ensemble, one finds that the stable ground state of the system is typically either `hot empty space'\,\footnote{In the present context, `hot empty space' refers to a space (with asymptotics appropriate to the model) that does not contain a black hole, but nevertheless has a finite period for the Euclidean time such that the boundary conditions of the ensemble are satisfied. Familiar examples include `hot flat space' \cite{Gross:1982cv} and thermal AdS \cite{Hawking:1982dh}.}, a regular black hole, or some superposition of the two, depending on the boundary conditions. Even when it is not the ground state a black hole may exist for certain boundary conditions as a local minimum of the free energy, stable against small thermodynamic fluctuations. We find that the inclusion of conical singularities in the ensemble does not change either of these statements. When the boundary conditions allow a thermodynamically stable black hole in the canonical ensemble, the addition of a small conical defect -- a conical singularity with deficit or surplus angle $|\alpha| \ll 1$ -- may result in a lower internal energy or a higher density of states. In other words, small conical defects may be energetically or entropically favored over their smooth counterparts in the conical ensemble. 
\begin{align}
E(\textrm{conical})-E(\textrm{smooth}) < 0 \textrm{\;is\;possible} \label{eq:intro1} \\
S(\textrm{conical})-S(\textrm{smooth}) > 0 \textrm{\;is\;possible} \label{eq:intro2}
\end{align}
These results are somewhat surprising because they seem to contradict our intuition that smooth black holes should be the most stable configurations. However, \eqref{eq:intro1} and \eqref{eq:intro2} are not possible simultaneously, and in fact the presence of a small conical defect always increases the free energy:
\eq{
F(\textrm{conical})-F(\textrm{smooth}) \sim \alpha^2 > 0\,, \,\,\, \textrm{\;for\;} \,\,|\alpha| \ll 1 ~.
}{eq:intro3}
We conclude that black holes that are stable against Gaussian fluctuations in the canonical ensemble are also stable against the nucleation of a small conical defect. This perturbative stability generalizes to a non-perturbative statement about the ground state of the system. With a few caveats, it appears that the ground state of the conical ensemble is always a smooth field configuration\,\footnote{For certain theories we find specific classes of solutions that are only marginally stable against the decay into conical defects, but these examples (the so-called `constant dilaton vacua') tend to suffer from other problems.}. In fact, the relationship between the ground state and the boundary conditions is the same as in the canonical ensemble. We provide general arguments as well as several examples demonstrating these features. 

In the semiclassical approximation the partition function is dominated by the ground state of the system, which is a smooth geometry. Corrections to this leading behavior from small, smooth fluctuations around the ground state are well-understood and can be evaluated using a variety of techniques. Since configurations with a conical singularity do not dominate the partition function, it is natural to ask how their contributions compare to the corrections from smooth fluctuations. The action is generally a complicated function of the fields, so contributions from conical singularities cannot be evaluated analytically except in special cases. However, it is possible to evaluate the contributions numerically, and in the semiclassical limit the results are approximated to high precision by relatively simple functions of the boundary conditions that define the ensemble. When the ground state is a black hole, we find that the contributions to the partition function from configurations with a conical singularity are comparable to the contributions from Gaussian fluctuations. This suggests that, even in the semiclassical approximation, non-smooth field configurations make important sub-leading contributions to thermodynamic quantities. 

It is important to point out that we view these contributions as logically distinct from the ``mass fluctuations'' that are already studied in the literature, see e.g.~\cite{Das:2001ic}, or from 1-loop quantum corrections derived by taking into account conical defects \cite{Solodukhin:1994yz}. Indeed, the reader may wonder whether the configurations we consider are not already included in those calculations. In the canonical ensemble the proper temperature is held fixed at the wall of the cavity, and hence black holes with a regular horizon are only present in the ensemble for isolated values of the mass (if they are present at all). Shifting the black hole mass away from these values necessarily introduces a conical singularity at the horizon, and this must be addressed before the role of these configurations in the ensemble can be understood. Thus, we assume that ``mass fluctuations'' in previous calculations like \cite{Das:2001ic} refer only to small, smooth fluctuations of the fields around a black hole background with a fixed horizon, as opposed to an actual variation in the mass of the black hole. On the other hand, 1-loop corrections depend explicitly on the precise matter content coupled to the gravitational theory, so they are not exclusively an intrinsic property of the black hole. See for instance \cite{Sen:2012dw}, Eqs.~(1.1), (1.2), and Refs.~therein. Thus, our approach of considering non-smooth geometric configurations that are on-shell everywhere except at the horizon is logically distinct form these two approaches. Nevertheless, as we will show, the leading corrections to quantities like the entropy take essentially the same form as they do in other approaches.

This paper is organized as follows. In section \ref{sec:1} we recall some basic results of  two-dimensional dilaton gravity and black hole thermodynamics in the canonical ensemble. In section \ref{sec:2} we study black holes with a conical defect and evaluate the conical ensemble partition function in the semiclassical approximation. In section \ref{sec:2a} we address general features of thermodynamical observables, discuss stability issues, and give approximate expressions for the contributions to the partition function. In section \ref{sec:3} we provide several explicit examples, and in section \ref{sec:4} we point to applications and open problems.

Before proceeding we point out a few important conventions, which are the same as \cite{Grumiller:2007ju} in most cases. Euclidean signature is employed throughout. Nevertheless, terms appropriate to Lorentzian signature such as `Minkowski' and `horizon' are used when the meaning is clear. We use natural units with $G_{d+1} = c = \hbar = k_{B} = 1$, and the dimensionless Newton's constant in two dimensions is set to $8\pi G_2 = 1$. These constants are restored in certain expressions, when necessary.

\section{Preliminaries}
\label{sec:1}

In this section we recapitulate some of the main results of \cite{Grumiller:2007ju}. A reader familiar with these results and the notation may skip this section. 
 
\subsection{Two-dimensional dilaton gravity action}

In this paper we study black hole (BH) thermodynamics in two-dimensional models of dilaton gravity. Dilaton gravity in two dimensions is conventionally described by the Euclidean action
\begin{align}\label{Action}
  \Gamma[g,X] = & \,\, - \frac{1}{2} \int_{\MM} \nts \nts \dd^{\,2}x \,\sqrt{g}\, \left( \,X\,R - U(X)\,\left(\nabla X\right)^2 - 2 \, V(X) \, \right) \\ \nonumber
& \,\, - \int_{\dM} \bns \dd x \,\sqrt{\gamma} \, X\,K 
+ \int_{\dM} \bns \dd x \,\sqrt{\gamma} \, \sqrt{w(X)e^{-Q(X)}} ~.
\end{align}
The dilaton $X$ is defined in terms of its coupling to the Ricci scalar, which takes the form $X R$. Different models are distinguished by the kinetic and potential functions $U(X)$ and $V(X)$, cf.~e.g.~\cite{Grumiller:2002nm,Grumiller:2006rc} for reviews. The bulk terms in the first line of \eqref{Action} are supplemented by boundary terms in the second line. The first boundary term is the analog of the Gibbons-Hawking-York surface integral \cite{York:1972sj, Gibbons:1976ue}, where $\gamma_{ab}$ is the induced metric on the boundary and $K$ is the trace of the extrinsic curvature. Including this term in the action ensures a Dirichlet boundary problem. The second boundary term is the holographic counterterm derived in \cite{Grumiller:2007ju}.  It ensures a well-defined variational principle, so that the first variation of the action vanishes on solutions of the equations of motion for all variations that preserve the boundary conditions\,\footnote{On non-compact spaces, the first variation of the action \eqref{Action} without the holographic counterterm vanishes only for field variations with compact support. It is worth mentioning that the specific combination of $w(X)$ and $Q(X)$ appearing in the counterterm is the supergravity pre-potential \cite{Grumiller:2007cj,Grumiller:2009dx}.}. The functions $w(X)$ and $Q(X)$, which depend on $U(X)$ and $V(X)$, are defined below.

\subsection{Equations of motion and all classical solutions}
\label{subsec:EOMandCS}

The equations of motion are obtained by extremizing the action \eqref{Action} with respect to small variations of the fields that preserve the boundary conditions. This yields
\begin{gather} \label{MetricEOM}
     U(X)\,\nabla_{\mu}X \nabla_{\nu}X - \frac{1}{2}\,g_{\mu\nu} U(X) (\nabla X)^2  
        - g_{\mu\nu} V(X) + \nabla_{\mu} \nabla_{\nu} X - g_{\mu\nu} \nabla^{2} X  = 0 \\ \label{XEOM}
     R + \partial_X U(X) (\nabla X)^2 + 2 \,U(X) \nabla^{2} X - 2 \,\partial_X V(X)  = 0 ~.
\end{gather}
Solutions of these equations always possess at least one Killing vector $\partial_{\tau}$ with orbits that are curves of constant $X$ \cite{Schmidt:1989ws, Banks:1990mk}. Fixing the gauge so that the metric is diagonal, the solutions take the form
\eq{
   X = X(r) \qquad \dd s^2 = \xi(X) \,\dd\tau^2 + \frac{1}{\xi(X)}\,\dd r^2
}{metric}
with
\begin{align}
  	\partial_r X = & \,\, e^{-Q(X)} \label{XPrimeDef} \\ \label{xiDef}
	\xi(X)  = & \,\, w(X) \, e^{Q(X)}\,\left( 1 - \frac{2\,M}{w(X)} \right)  ~.
\end{align} 
The solutions depend on an integration constant $M$, as well as two model-dependent functions $Q(X)$ and $w(X)$ that are given by integrals of $U(X)$ and $V(X)$
\begin{eqnarray}\label{QDef}
	Q(X) & \defeq & Q_0 \, + \int^{X} \bns \dd\tilde{X} \, U(\tilde{X}) \\ \label{wDef}
        w(X) &  \defeq & w_0 -2 \, \int^{X} \bns \dd\tilde{X} \, V(\tilde{X}) \, e^{Q(\tilde{X})} ~.
\end{eqnarray} 
The integrals are evaluated at $X$, with constants of integration $Q_0$ and $w_0$. Notice that $w_0$ and $M$ contribute to $\xi(X)$ in the same manner; together they represent a single parameter that has been partially incorporated into the definition of $w(X)$. By definition they transform as $w_0 \to e^{\Delta Q_0} w_0$ and $M \to e^{\Delta Q_0} M$ under the shift $Q_0 \to Q_0 + \Delta Q_0$. This ensures that the functions \eqref{XPrimeDef} and \eqref{xiDef} transform homogeneously, allowing $Q_0$ to be absorbed into a rescaling of the coordinates\,\footnote{Note that the counterterm in the action \eqref{Action} depends on $w_0$ but not on $Q_0$.}. Therefore, the solution depends on a single constant $w_0 + M$. 
With an appropriate choice of $w_0$ we can restrict $M$ to take values in the range $M \geq 0$ for physical solutions. As evident from \eqref{metric} the norm of the Killing vector $\partial_{\tau}$ is $\sqrt{\xi(X)}$. If it vanishes we encounter a Killing horizon. Solutions with $M > 0$, which exhibit horizons, will be referred to as BHs.

If the function $V(X)$ happens to have a zero at $X_\ts{CDV}$ then there is a second, inequivalent family of solutions that also have the form \eqref{metric}. The dilaton and metric for these solutions are given by
\begin{eqnarray}
  	X & = & X_\ts{CDV} \label{XCDV} \\ \label{xiCDV}
	\xi & = & \hat c + \hat a\, r - V'(X_\ts{CDV})\,r^2  ~,
\end{eqnarray} 
where $\hat{c}$ and $\hat{a}$ are arbitrary constants. In most applications these solutions, which are characterized by a constant dilaton and Ricci scalar, are not relevant. We will generally ignore them, so references to ``generic solutions'' or ``all solutions'' should be understood to mean the solutions \eqref{XPrimeDef} and \eqref{xiDef}, parametrized by the mass $M$.

\subsection{Smooth Black Holes}

In the models we consider, solutions have a non-negative metric function $\xi(X)$ over a semi-infinite interval 
\begin{equation}\label{Interval}
X_\text{min} \leq X < \infty ~,
\end{equation}
with the lower end of this interval corresponding to either the origin or a horizon, and the upper end corresponding to the asymptotic region of the space-time. At the upper end of the interval the function $w(X)$ generally diverges 
\begin{equation}\label{wAsymptotic}
   \lim_{X \to \infty} w(X) \to  \infty~,
\end{equation} 
so the asymptotic behavior of the metric is characterized by the solution with $M=0$. 
If the metric function is strictly positive then the lower end of the interval \eqref{Interval} is just the value of the dilaton at the origin. But if $\xi(X)$ vanishes for some value of the dilaton, $X=X_h$, then the lower bound is a Killing horizon. Assuming that the function $e^{Q(X)}$ is non-zero for finite values of $X$, the location of the horizon is related to the parameter $M$ by
\begin{equation}\label{Horizon}
	w(X_h) = 2M  ~. 
\end{equation}
If this condition admits multiple solutions then $X_h$ is always taken to be the outermost horizon, so that $w(X) > 2M$ for $X > X_h$.

For a field configuration to extremize the action it should satisfy the equations of motion at all points, and this imposes certain differentiability conditions on solutions. In particular, for solutions with $M \neq 0$ the horizon must be a regular point of the geometry. This fixes the periodicity $\tau \sim \tau + \beta$ of the Euclidean time, 
which is given by \cite{Gegenberg:1994pv}
\begin{equation}\label{beta}
   \beta = \left. \frac{4\pi}{\partial_r \xi}\, \right|_{r_h}  
	= \left. \frac{4\pi}{w'(X)} \,\right|_{X_h} ~.
\end{equation}
The inverse periodicity is related to the surface gravity of the BH by $2\pi \beta^{-1} = \kappa$. In an asymptotically flat space-time $\beta^{-1}$ is also the temperature measured by an observer at infinity, so we denote this quantity by $T$
\begin{equation}\label{T}
  T = \frac{1}{\beta} =  \left. \frac{w'(X)}{4\,\pi}  \right|_{X_h} ~.
\end{equation}
This slight abuse of notation should not be confused with the proper local temperature $T(X)$, which is related to $\beta^{-1}$ by a dilaton-dependent `Tolman factor'\cite{Tolman:1934}
\begin{equation}\label{Tc}
  T(X) =  \frac{1}{\beta\,\sqrt{\xi(X)}} ~.
\end{equation}
The proper temperature at infinity coincides with \eqref{T} only if $\xi(X) \to 1$ as $X \to \infty$.

A solution with $M=0$ has no horizon, and is therefore regular everywhere without having to place any conditions on the period of the Euclidean time. However, as we will see below, the boundary conditions of the canonical ensemble determine a unique, non-zero value for the otherwise arbitrary period. We will therefore refer to this solution, which does not contain a black hole but has a non-zero temperature, as `Hot Empty Space' (HES).

\subsection{Thermodynamics in the Canonical Ensemble}

To describe a consistent BH thermodynamics we must specify an ensemble and construct the appropriate partition function. Motivated by York's path integral formulation of the canonical ensemble \cite{York:1986it}, we introduce an upper bound $X_c$ on the interval \eqref{Interval}. This constrains the dilaton to a `cavity' $X \leq X_c$ whose wall is the dilaton isosurface $X=X_c$. Quantities evaluated at the wall or with an explicit dependence on $X_c$ will carry a subscript `$c$'.

Boundary conditions for the canonical ensemble are imposed by coupling the system to a thermal reservoir, which fixes a dilaton charge $D_c$ and the proper local temperature $T_c$ at the wall of the cavity\,\footnote{There is no unique dilaton charge in two dimensions: given any function of $X$, one can construct a current that yields that function as its conserved charge. For simplicity, we take the dilaton charge at the wall to be $D_c = X_c$, and refer to this boundary condition henceforth as fixing $X_c$. A detailed discussion can be found in \cite{Gibbons:1992rh}, or in section 3.1 of \cite{Grumiller:2007ju}.}. It is convenient to think of the boundary condition on the temperature as fixing the proper local period of the Euclidean time, which is just the inverse $\beta_c = T_{c}^{\,-1}$ of the proper local temperature. The proper local period is related to the period $\tau \sim \tau + \beta$ by 
\begin{gather}\label{periodBC}
	\beta_c := \beta\,\sqrt{\xi_{c}} ~.
\end{gather}
When combined with the smoothness condition \eqref{T} this becomes
\begin{gather}\label{SmoothPeriodBC}
	\beta_c = \frac{4\pi}{w'(X_h)}\,\sqrt{\xi_{c}}~.
\end{gather}
This model-dependent (and often complicated) equation, which may or may not have solutions $M >0$, determines whether there are smooth black holes in the ensemble for given boundary conditions $\beta_c$ and $X_c$. Not all solutions of this equation are relevant: the upper bound on the dilaton implies that only solutions with $X_{h}(M) < X_c$ `fit' inside the cavity. Thus, any solutions $M$ of \eqref{SmoothPeriodBC} that lie in the range $0 \leq M < M^\ts{max}$ are elements of the canonical ensemble, where 
\begin{gather}
	M^\ts{max} = \frac{1}{2}\,w(X_c)
\end{gather}
corresponds to a black hole with horizon located at the wall of the cavity. One solution that almost always appears in the canonical ensemble is HES ($M=0$) with period fixed by the boundary condition \eqref{periodBC}
\begin{gather}\label{HESperiod}
	\beta_\ts{HES} = \frac{\beta_c}{\sqrt{e^{Q_c}\,w_c}} ~.
\end{gather}
In most models the HES solution dominates the ensemble for at least some range of boundary conditions \cite{York:1986it, Hawking:1982dh}.

With these boundary conditions the partition function of the canonical ensemble is given by the Euclidean path integral
\begin{equation}\label{PartitionFunction}
  \ZZ = \int_{_{X_c, T_c}} \bns \bns \CurlD g \CurlD X \, \exp\left(-\Ga[g,X] \right) ~.
\end{equation}
For now we will take the path integral to include all smooth spaces $(\MM,g)$ and dilaton configurations $X$ that satisfy the boundary conditions, but we will relax the smoothness requirement in the next section. In the semi-classical limit the dominant contribution to the Euclidean path integral comes from the minimum of the action. The minimum is of course a stationary point of the action -- either a black hole with $M > 0$ satisfying \eqref{SmoothPeriodBC} (if such a solution exists), or HES with period \eqref{HESperiod}. So the action for smooth field configurations near the minimum can be written as
\begin{align}\label{Min}
  \Ga[g_\ts{min} + \delta g, X_\ts{min} + \delta X] = & \,\, \Ga[g_\ts{min},X_\ts{min}] + \de \Ga[g_\ts{min}, X_\ts{min}; \delta g, \delta X] \\ \nonumber
  & \,\, + \frac12 \,\de^2\Ga[g_\ts{min}, X_\ts{min}; \delta g, \delta X] + \ldots	
\end{align}
where $\de \Ga$ and $\de^2 \Ga$ are the linear and quadratic terms in the Taylor expansion. In \cite{Grumiller:2007ju} it was shown that the leading term is finite and the linear term vanishes for all field variations $\delta g$, $\delta X$ consistent with the boundary conditions. Since the solution is assumed to be a minimum (as opposed to a saddle point) the quadratic term is positive definite and the semi-classical approximation of the path integral is given by
\begin{equation}\label{ApproxPF}
 \ZZ \approx \exp\left(-\Ga[g_{\ts{min}},X_{\ts{min}}]\right) \times (\text{Quadratic Corrections})~,
\end{equation}
where the second factor comes from performing the (Gaussian) integral over the quadratic terms in \eqref{Min}.

The partition function exhibits qualitatively different behavior depending on whether the minimum of the action is HES or a black hole. For a particular model the ground state can be readily determined from the values of the boundary conditions: one simply identifies the solutions of the equations of motion that belong to the ensemble, and then determines which solution has the smallest action for the given values $\beta_c$ and $X_c$. Evaluating the action \eqref{Action} for the solutions \eqref{XPrimeDef} and \eqref{xiDef} gives
\eq{
\Ga_c(M)=\be_c \sqrt{w_ce^{-Q_c}}\left(1-\sqrt{1-\frac{2M}{w_c}}\,\right)-2\pi\,X_h(M) ~,
}{eq:gammaM}
which is bounded below for finite $X_c$. 
For HES with $M = 0$ this becomes
\begin{gather}\label{HESaction}
	\Ga_c(0) = -2\pi\,X_0 ~,
\end{gather}
where $X_0$ is the value of the dilaton at the origin (in most of the models we consider, $X_0 = 0$). If \eqref{HESaction} is less than \eqref{eq:gammaM} for all relevant solutions of \eqref{SmoothPeriodBC}, then HES is the ground state of the ensemble. On the other hand, if there is a solution of \eqref{SmoothPeriodBC} such that \eqref{eq:gammaM} is less than \eqref{HESaction}, then the ground state of the ensemble is a black hole. If the values of $\beta_c$ and $X_c$ are changed then the ground state of the ensemble may change as well, in which case the system will undergo a phase transition involving the nucleation of the new ground state -- either a stable black hole or HES. Assuming that there is a single minimum of the action\,\footnote{For special values of the boundary conditions there may be multiple values of $M$ that minimize the action. For instance, it may be possible to tune $\beta_c$ and $X_c$ so that HES and a black hole both have the same action.}, the dominant semiclassical contribution to the free energy $F_c = - T_c \,\ln{\ZZ}$ is given by
\begin{equation}\label{ZFRelation}
  F_c (T_c, X_c) \simeq T_c \, \Gamma_c(M) =  \sqrt{w_c\,e^{-Q_c}}\,\left(1 - \sqrt{1-\frac{2\,M}{w_c}}\right)- 2\pi\,X_{h}(M) \, T_c ~.
\end{equation}
From this result it is possible to derive all thermodynamical properties of interest, like the entropy and internal energy
\begin{align}
	S := & \, - \left(\frac{\partial F_{c}}{\partial T_c}\right)_{X_c} = 2\pi X_{h}(M) \label{eq:entropy} \\
	E_{c} := & \, F_{c} + T_{c}\,S = \sqrt{w_{c}\,e^{-Q_{c}}}\,\left(1 - \sqrt{1-\frac{2\,M}{w_{c}}}\,\right) ~.
\end{align}
A comprehensive discussion of thermodynamical properties is provided in \cite{Grumiller:2007ju}.

A key assumption in the derivation of \eqref{ApproxPF} is that the quadratic term in \eqref{Min} must be positive definite. This is just the thermodynamic stability condition that the ground state have positive specific heat at constant dilaton charge. The specific heat at constant $X_c$ is  
\begin{gather}
	C_{c} \defeq \frac{\partial E_c}{\partial T_c}\bigg|_{X_c} = T_c \, \frac{\partial S}{\partial T_c}\bigg|_{X_c}
	\qquad \textrm{with} \quad E_c = F_c + T_c \, S ~,
\end{gather}
which yields
\begin{gather}\label{CD}
	C_{c} = \frac{4\pi\,w_{h}'\,(w_c - 2M)}{2\,w_{h}''\,(w_c - 2M) + (w_{h}')^2} ~. 
\end{gather}
Thus, given boundary conditions $X_c$ and $T_c$, a canonical ensemble dominated by a black hole exists if the minimum of the action is a solution $0 < M < M^\ts{max}$ of \eqref{SmoothPeriodBC}, and
\begin{gather}\label{CDinequality}
	w_{h}'' > - \frac{(w_{h}')^2}{2\,(w_c - 2 M)}
\end{gather}
so that the specific heat \eqref{CD} is positive. An important point is that for some theories this inequality can only be realized for finite $X_c$. A theory whose boundary conditions and solutions have $w''(X_h) < 0$ will not have positive specific heat as $X_c \to \infty$, since $w_c \to \infty$ in this limit. In that case a finite cavity is required for the existence of the canonical ensemble. The classic example of this phenomenon is the Schwarzschild black hole in a theory with asymptotically flat boundary conditions \cite{York:1986it}. On the other hand, the right-hand side of \eqref{CDinequality} is strictly negative, so a finite cavity is not required for a theory with boundary conditions and solutions such that $w''(X_h) > 0$.  In that case the specific heat remains positive as $X_c \to \infty$. If the action of the black hole is less than the action for HES in this limit, then there is a canonical ensemble with black hole ground state that does not require an external thermal reservoir. The expression \eqref{CD} for the specific heat will play an important role in later sections. 

This concludes our review of thermodynamics in the Euclidean path integral formalism for two-dimensional dilaton gravity. In the rest of the paper we weaken the assumption of differentiability by considering continuous metrics that no longer satisfy the condition \eqref{beta}.

\section{Black holes with conical defect}
\label{sec:2}

Let us now reconsider the Euclidean partition function \eqref{PartitionFunction}. In the previous section we included contributions from smooth field configurations that satisfy the boundary conditions. Then the leading term in the semi-classical approximation of the canonical partition function is
\eq{
\ZZ \approx \exp\Big(-\Gamma_{c}(M)\Big) ~,
}{eq:con101}
with $M$ being the mass of the smooth solution that minimizes the Euclidean action and $\Gamma_c$ the on-shell action \eqref{eq:gammaM}. We assume as before that the absolute minimum of the action occurs for a single value of $M$; otherwise \eqref{eq:con101} contains a sum over all values of $M$ that minimize the action. We may rewrite \eqref{eq:con101} as 
\eq{
\ZZ \sim \int\limits_{0}^{M^\ts{max}} \nts \dd\hat{M} \,\de(\hat{M} - M)\,\exp\Big(-\Gamma_c(\hat{M})\Big) 
}{eq:con102} 
where $0 \leq \hat{M} < M^\ts{max}$ runs over the physically allowed values of the mass -- subject to the condition $X_{h}(\hat{M}) < X_c$ -- and the delta-function picks out the value $\hat{M} = M$ of the smooth solution that minimizes the action.  

Now suppose that we enlarge the class of field configurations that contribute to $\ZZ$ by relaxing the assumption of smoothness. Instead of imposing the condition \eqref{beta} for the period, we allow for metrics that are continuous but not differentiable at the center of the cavity. In the semiclassical limit, the largest contributions to the partition function from this sector are expected to come from configurations that `almost' meet the conditions for stationary points of the action: they satisfy the equations of motion at all points except for the horizon, where there is a conical singularity. Assuming the action is well-defined for these backgrounds, the contributions to $\ZZ$ take the form \eqref{eq:con102} without the delta-function\,\footnote{In later sections we will refine the measure of this integral, expressing the contributions to $\ZZ$ as an integral over internal energy $\dd E_{c}$ with the weight given by $\exp(-\beta_{c}\,F_{c}(\hat{M})) = \exp(S)\,\exp(-\beta_{c}\,E_{c})$.}
\eq{
\ZZ \approx \int\limits_{0}^{M^\ts{max}} \nts \dd\hat{M} \,\exp\Big(-\Gamma_c(\hat{M})\Big) ~.
}{eq:con102b} 
In the rest of this section we will consider the properties of the `conical defect' black holes that contribute to this integral, and evaluate the action $\Gamma_{c}(\hat{M})$ that appears in the exponent.

\subsection{Classical black hole solutions with conical defect}

Black hole field configurations that satisfy the boundary conditions but \emph{not} the smoothness condition \eqref{beta} have the same general form as before
\eq{
   X = X(r)\,, \qquad \dd s^2 = \xi(\hat{M}, X) \,\dd\tau^2 + \frac{1}{\xi(\hat{M},X)}\,\dd r^2\,,
}{metricc}
with period $\tau \sim \tau + \tilde{\beta}$, and the functions $X$ and $\xi$ given by
\begin{eqnarray}
  	\partial_r X & = & e^{-Q(X)} \label{XPrimeDefc} \\ \label{xiDefc}
	\xi(\hat{M},X) & = & w(X) \, e^{Q(X)}\,\left( 1 - \frac{2\,\hat M}{w(X)} \right) ~.
\end{eqnarray} 
The location of the horizon, $X_{h}(\hat{M}) < X_c$, is determined as before: 
\eq{
X_{h}(\hat{M}) = w^{-1}(2\hat M) ~.
}{eq:con202}
We will sometimes denote a function's dependence on $\hat{M}$ (as opposed to the values $M$ that satisfy \eqref{SmoothPeriodBC}, which typically comprise a discrete set) with a `hat', abbreviating $X_{h}(\hat{M})$ as $\hat{X}_{h}$ and $\xi(\hat{M},X)$ as $\hat{\xi}(X)$. 

By assumption, most values of the parameter $\hat{M}$ do not correspond to smooth black holes -- the condition \eqref{SmoothPeriodBC} is not satisfied. This means that the periodicity of the Euclidean time is not equal to the period required for regularity at the horizon. Instead, the period $\tau\sim\tau+\tilde\be$ is determined by the boundary condition \eqref{periodBC} and the parameter $\hat M$,
\eq{
	\tilde\be = \frac{\be_{c}}{\sqrt{\hat{\xi}_{c}}} ~.
}{eq:con201}
In other words, the period $\tilde\be$ does not agree with $\hat\be$, defined as
\begin{gather}
	\hat{\be} := \frac{4\pi}{w'(X)}\bigg|_{\hat{X}_{h}} ~.
\end{gather}
As a result, these spaces exhibit a a conical singularity. The deficit (or surplus) angle $\deficit$ associated with the defect is
\eq{
\deficit \,:=\, 2\pi \,\frac{\hat\be-\tilde\be}{\hat\be} \, = \, 2\pi\,\left(1 - \frac{\be_c}{\hat\be_c}\,\right) ~,
}{eq:con203}
where $\hat\be_c := \hat\be\,\sqrt{\hat\xi_c}$\,. If $\hat\be > \tilde\be$ then $\alpha$ is positive, and it represents a true deficit in the period of Euclidean time. Otherwise, if $\alpha$ is negative there is a surplus in the period. For convenience we will always refer to $\deficit$ as the `deficit angle', though it may be positive or negative.

An important distinction between spaces with a conical defect and smooth solutions of the equations of motion is that $\hat{M}$ is a continuous parameter that is independent of the boundary conditions $\beta_c$ and $X_c$. The only conditions on $\hat{M}$ are that it should lie in the range associated with physical solutions, $\hat{M} \geq 0$, and the horizon should fit inside the cavity, $X_{h}(\hat{M}) < X_c$. The discrete set of masses $M$ that correspond to smooth solutions are determined by the condition \eqref{SmoothPeriodBC}, which implies a (potentially complicated) dependence on $\beta_c$ and $X_c$. 

\begin{figure}
  \centering
	\includegraphics{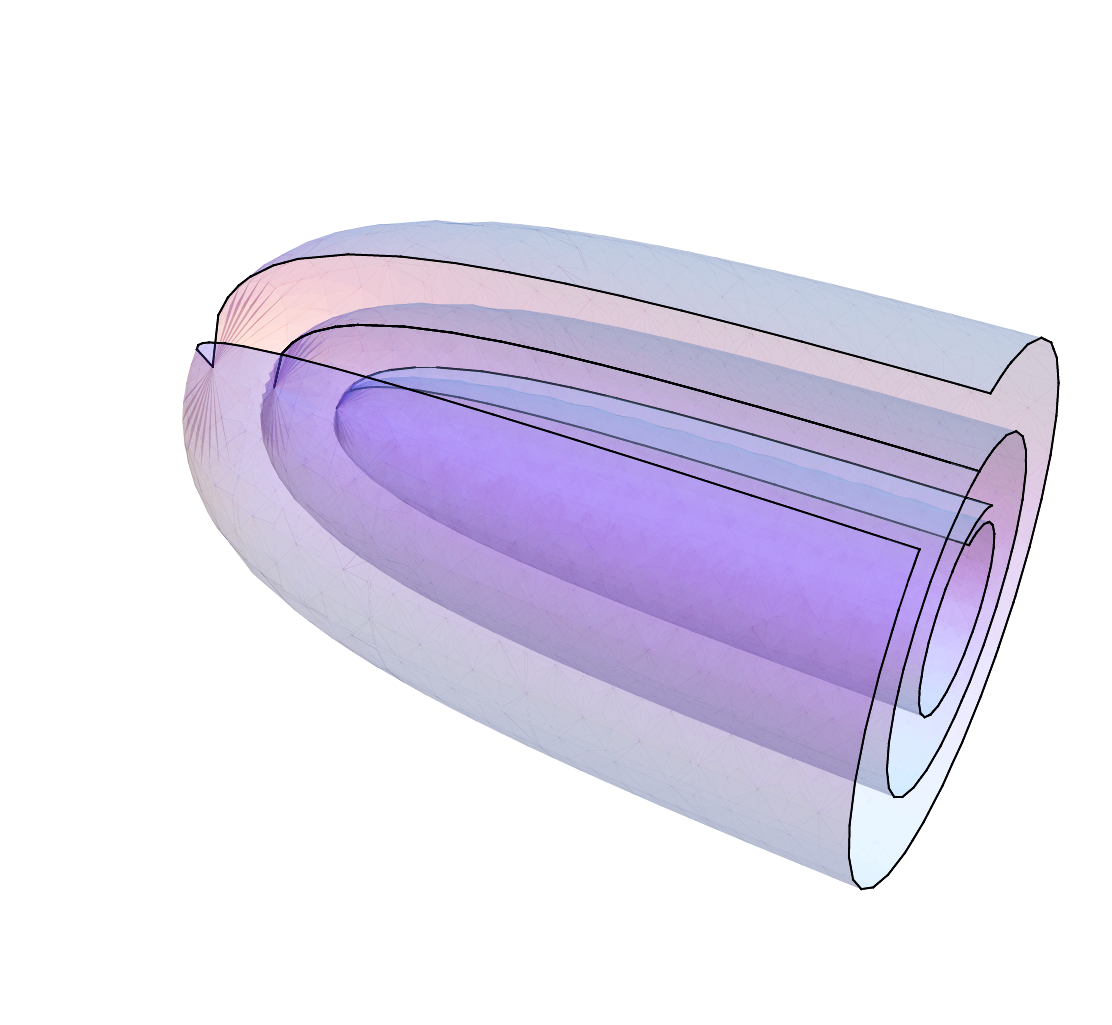}	
  \caption{Euclidean BH geometries with positive (outer), vanishing (middle), and negative (inner) deficit angles.}
\end{figure}

\subsection{Euclidean action}

In the presence of a conical singularity the action must be evaluated carefully, taking into account the behavior of the metric at the horizon. This is accomplished using a smoothing procedure that regulates the defect \cite{Farhi:1989yr,Hayward:1993my,Brill:1994mb,Fursaev:1995ef}. When the action is evaluated for a black hole with conical singularity the result is independent of the details of the smoothing. This suggests that the action \eqref{Action}, without any modifications, is appropriate for weighting contributions to the partition function from these spaces, as in \eqref{eq:con102b}.

For our purposes, a conical singularity at a point $p$ on the interior of $\MM$ can be thought of as introducing a delta-function in the curvature, in the sense that the integral of the Ricci scalar over $\MM$ is
\begin{gather}
	\int_{\MM} \nts \dd^{2}x \,\sqrt{g} \, R = 2 \,\alpha + \int_{\MM/p} \bns\nts \dd^{2}x \,\sqrt{g}\, R ~,
\end{gather}
where $\alpha$ is the deficit angle and $\MM/p$ is the manifold $\MM$ with the point $p$ removed. 
The spaces described in the previous section have a conical defect at the horizon $\hat{X}_h$, so we can write the action as the usual functional on $\hat{\MM}$ -- the manifold $\MM$ with the singular point removed -- plus the contribution from the defect
\begin{align}\label{Actionc}
  \Gamma[g,X] = & \,\, - \frac{1}{2}\,\int_{\hat\MM} \nts \dd^{\,2}x \,\sqrt{g}\, \left[ X\,R - U(X)\,\left(\nabla X\right)^2 - 2 \, V(X) \right] \\ \nonumber
		   & \,\, - \int_{\dM} \bns \dd x \,\sqrt{\gamma} \, X\,K + \int_{\dM} \bns \dd x \,\sqrt{\gamma} \, \sqrt{w(X)e^{-Q(X)}} - \hat{X}_h\,\deficit ~.
\end{align}
For $V(X)=0$ and constant dilaton the resulting functional of the metric is, up to an overall factor of $-2\pi X$, the Gauss-Bonnet formula for a compact Euclidean manifold $\MM$ with boundary $\dM$ and a deficit angle $\deficit$.

With the results above the on-shell action for a black hole with a conical singularity at the horizon is
\eq{
\Ga_c(\hat M)=\be_{c}\,\sqrt{w_{c}\,e^{-Q_c}}\,\Bigg(\,1-\sqrt{1-\frac{2\hat M}{w_c}}\,\Bigg)-2\pi\,X_{h}(\hat{M}) ~.
}{eq:gammac}
The free energy, which encodes all thermodynamical properties of interest, is then given by 
\eq{
F_{c}(\hat{M})\, = \, T_c \, \Ga_{c}(\hat{M}) \, = \, \sqrt{w_{c}\,e^{-Q_c}}\,\Bigg(\,1-\sqrt{1-\frac{2\hat M}{w_c}}\,\Bigg)-2\pi\,X_{h}(\hat{M})\, T_c ~.
}{eq:F}
When considering the role of black holes with a conical singularity in the ensemble it is useful to compare \eqref{eq:gammac} to the action for a smooth solution of the equations of motion. The difference between the actions is given by
\begin{align}\label{eq:deGa}
	\De\Ga := &\,  \Ga_c(\hat M)-\Ga_c(M) \\ 
			= &\, \be_c \, \sqrt{w_{c}\,e^{-Q_c}}\,\left(\,\sqrt{1 - \frac{2\,M \vphantom{\hat{M}}}{w_c}} - \sqrt{1-\frac{2\,\hat{M}}{w_{c}}} \,\right) + 2 \pi \big(X_{h}(M) - X_{h}(\hat{M})\big)
\end{align}
This result will be useful in the next section.

\section{Thermodynamics and stability}\label{sec:2a}

We investigate now the thermodynamical properties of field configurations with a conical defect, and compare their role in the conical ensemble to that of smooth solutions. For now we assume that the ensemble contains among the smooth field configurations a black hole with mass $M$, horizon $X_{h}(M)$, and positive specific heat. Thus, the black hole is at least a local minimum of the action among smooth spaces, though it need not be the absolute minimum of the action.  It is useful to define a quantity $\de$ that relates the dilaton at the conical defect to the dilaton at the horizon of the smooth solution 
\eq{
X_{h}(\hat{M}) = X_{h}(M) +\de ~.
}{eq:con12}
After deriving expressions for the entropy and internal energy, we consider the case
\eq{
|\de| \ll X_h ~,
}{eq:pert}
where the thermodynamic properties of conical defects can be analyzed perturbatively. Later in this section we derive results that are valid non-perturbatively, in particular concerning the stability of smooth configurations and the ground state of the ensemble. 

\subsection{Entropy and Internal Energy}

In two-dimensional dilaton gravity the entropy of smooth black holes takes the universal form \eqref{eq:entropy}. This result, which is independent of both the details of the theory and the size of the cavity, generalizes to black holes with a conical singularity
\begin{gather}\label{eq:con20}
	S(\hat{M}) := - \left(\frac{\partial F_{c}(\hat{M})}{\partial T_c}\right)_{X_c} = \,2\pi\,X_{h}(\hat{M}) ~.
\end{gather}
In terms of the parameter $\delta$ this becomes
\begin{gather}\label{Sdelta}
	S(\hat{M}) = S(M) + 2\pi\,\delta ~,
\end{gather}
and we see that the entropy may either be greater than or less than the entropy of the smooth black hole, depending on the sign of $\delta$.

The internal energy is related to the entropy, temperature, and free energy by $E_{c} = F_{c} + T_{c}\,S$. The subscript `c' is retained here to emphasize that, unlike the entropy, the internal energy depends explicitly on the size of the cavity. Applying the results for the entropy \eqref{eq:con20} and the free energy \eqref{eq:F} gives
\begin{gather}\label{InternalEnergy}
	E_{c}(\hat{M}) = \sqrt{w_{c}\,e^{-Q_{c}}}\,\left( 1 - \sqrt{1 - \frac{2\,\hat{M}}{w_c}}\,\right) = \sqrt{w_{c}\,e^{-Q_{c}}}\,\left( 1 - \sqrt{1 - \frac{w(X_{h} + \de)}{w_c}}\,\right) ~.
\end{gather}
Like the entropy, the internal energy of a black hole with a conical singularity may be either greater than or less than that of the smooth black hole.

\subsection{Perturbative stability}

Based on the results above, a black hole with a conical singularity can have higher entropy or lower internal energy than a smooth black hole. However, to determine whether these black holes are favored in the conical ensemble one must consider the free energy. We will first address this for small $\de$; i.e., $|\de| \ll X_h$. In this limit the deficit angle is
\begin{gather}\label{SmallDefect}
	\alpha = - \frac{4\pi^{2}}{C_{c}}\,\de + \mathcal{O}(\de^{2}) ~,
\end{gather}
so positive $\de$ corresponds to a surplus, and negative $\de$ represents a deficit. 

The expression \eqref{eq:con20} for the entropy is linear in $X_{h}(\hat{M})$, so no expansions are needed when $\de$ is small. On the other hand, the internal energy is a non-linear and potentially complicated function of $X_{h}(\hat{M})$. Expanding \eqref{InternalEnergy} for small $\de$ gives
\begin{gather}\label{Edelta}
	E_{c}(\hat{M}) \simeq E_{c}(M) + 2\pi\,T_c\,\de + \frac{2\pi^{2}\,T_c}{C_{c}}\,\de^{2} + \mathcal{O}(\de^{3}) ~.
\end{gather}
So the internal energy of the conical defect may be greater or less than the internal energy of the smooth black hole, and at leading order this is controlled by the sign of $\de$. Notice, however, that the term at order $\mathcal{O}(\de^{2})$ is strictly positive. This is crucially important when we consider the free energy $F_{c}(\hat{M})$. Using \eqref{Sdelta} and the expansion \eqref{Edelta} we obtain
\begin{align}
	F_{c}(\hat{M}) = E_{c}(\hat{M}) - T_{c}\,S(\hat{M}) \simeq F_{c}(M) + \delta^{2}\,\frac{2\pi^{2} T_{c}}{C_{c}} + \mathcal{O}(\de^3) ~.
\end{align}
Expressed in terms of the deficit angle this is
\begin{gather}
	F_{c}(\hat{M}) - F_{c}(M) \,\simeq \frac{C_{c}}{8 \pi^{2}}\,\alpha^{2} \quad \quad \alpha \ll 1 ~. 
\end{gather}
Thus, the free energy of a smooth black hole with $C_c > 0$ is always smaller than the free energy of a nearby ($|\de| \ll X_{h}(M)$) black hole with conical singularity. In terms of the internal energy and entropy this implies
\begin{gather}
	E_{c}(\hat{M}) - E_{c}(M) \,\ge\, T_c\, \left( S(\hat{M}) - S(M) \right) ~.
\end{gather}
In other words, the presence of a small conical defect can lower the internal energy compared to a smooth black hole, but the corresponding decrease in the density of states prevents the ensemble from favoring such configurations. Likewise, a conical defect can have a larger entropy than a smooth black hole, but the cost in internal energy is too high for these configurations to be favored by the ensemble.

\subsection{Non-perturbative stability}

The previous section considered black holes with a small conical defect. In this section the assumption $|\de| \ll X_{h}$ is dropped, which means that quantities like $\De \Ga_c$ cannot be evaluated perturbatively. Nevertheless, it is still possible to show that the minimum of the action does not have a conical defect. If the absolute minimum of the action among smooth spaces is a solution with mass $M$, then $\Ga_{c}(\hat{M}) > \Ga_{c}(M)$ for any $\hat{M}$ with a conical singularity. The ensemble always has a smooth ground state that is stable against decay into a space with conical defect.

First let us illustrate our reasoning with a simple class of examples: theories that allow the $X_c \to \infty$ limit. The existence of the ensemble in this limit is addressed in the next section; for now let us assume that we are working with a model where taking $X_c \to \infty$ is allowed. Then as the cavity wall is removed \eqref{eq:deGa} becomes 
\eq{
\lim_{X_c\to\infty}\De\Ga = 2\pi\, \frac{\de}{w^{\prime}(X_{h})} \left(\frac{w(X_h+\de)-w(X_h)}{\de}-w^{\prime}(X_{h})\right) ~.
}{eq:con21}
The condition $\De\Ga\geq 0$ becomes a convexity condition on the function $w$. An even stronger statement is obtained by considering extrema of $\De\Ga$. For very large $X_c$ the condition
\eq{
\frac{\dd\De\Ga}{\dd\,\de}=0
}{eq:con15} 
simplifies to
\eq{
w^\prime(X_h) = w^\prime(X_h+\de)
}{eq:con16}
But this implies that any extremum of $\De\Ga$ has to have the same periodicity as the configuration without conical defect. While more than one such extremum may exist (cf.~the discussion about how to extract the mass $M$ from $T_c$ and $X_c$ in \cite{Grumiller:2007ju}), none of them exhibits a conical singularity. 

For finite values of $X_c$ it is easier to work directly with the action \eqref{eq:gammac} for a space with conical singularity. Extremizing the action with respect to $\hat{M}$ gives
\begin{gather}
	\beta_{c} = \frac{4\pi}{w'(\hat{X}_{h})}\,\sqrt{\hat{\xi}_{c}} ~,
\end{gather}
which means that extrema occur at precisely those values of $\hat{M}$ that correspond to smooth solutions of the equations of motion. Of course, it is possible that these extrema are local, and the absolute minimum of the action occurs at one of the endpoints of the interval $0 \leq \hat{M} \leq M^\ts{max}$. Indeed, in most models there is a range of boundary conditions where the minimum of the action occurs at the lower limit. But $\hat{M} = 0$ is just HES, which is a smooth solution of the equations of motion. The other possibility -- that the absolute minimum of the action occurs at $M^\ts{max} = w(X_c)/2$ -- can be ruled out quite generally. Consider the derivative $\partial \Gamma_{c}(\hat{M})/\partial \hat{M}$ as $\hat{M} \to M^\ts{max}$ from below 
\begin{gather}\label{extrema1}
	\frac{\partial\,\Gamma_{c}(\hat{M})}{\partial\,\hat{M}} = \frac{\beta_{c}}{\sqrt{\hat{\xi}_{c}}} - \frac{4\pi}{w'(\hat{X}_{h})} = 0 ~.
\end{gather}
For non-zero $\beta_c$ the first term is positive and diverges as $(M^\ts{max} - \hat{M})^{-1/2}$. Unless $w'(X)$ happens to have a zero at $X_c$, the first term dominates and the action is \emph{increasing} as $\hat{M}$ approaches $M^\ts{max}$. We conclude that the action is always minimized by a smooth solution of the equations of motion: either HES or a smooth black hole.

Note that it is possible for a conical singularity to have a smaller action than a smooth black hole, as long as that black hole is not the ground state of the ensemble. This includes black holes that are thermodynamically stable ($C_{c}>0$), but only a local minimum of the action. In that case there will necessarily be conical singularities close to the ground state that have a smaller action than any local minimum. But a smooth black hole (or HES, for that matter) will never tunnel quantum mechanically to a final configuration with a defect, because the ground state of the ensemble is necessarily smooth.

\subsection{Constant dilaton vacua}

The discussion up to this point has involved generic solutions of the equations of motion, but neglected the constant dilaton vacua (CDV) \eqref{XCDV}-\eqref{xiCDV} that may exist for some dilaton gravity models. These isolated solutions occupy a different superselection sector of the theory, so there is no perturbative channel for a BH -- with or without a deficit angle -- to decay into a CDV. However, in cases where the boundary conditions happen to coincide with a zero of the dilaton potential, $V(X_c) = 0$, tunneling between the two types of solutions is possible. A detailed discussion can be found in \cite{Grumiller:2007ju}.

Since we have extended the class of BH solutions to configurations with a deficit angle, it is appropriate to do the same for CDVs. Here we discuss these solutions and evaluate their free energy. The on-shell action can be calculated using \eqref{Actionc}, which gives
\eq{
\hat{\Ga}_{CDV} = -2\pi X_0 + \tilde{\be}\,\sqrt{\hat{\xi}_c} \, \sqrt{e^{-Q(X_0)}w(X_0)} ~.
}{eq:con30}
(If we drop the assumption that spacetime is topologically a disk the first term in \eqref{eq:con30} is multiplied by the Euler characteristic of the manifold.)
The difference between this action and the action for a smooth CDV solution is
\eq{
\De\Ga_{CDV} = \left( \tilde{\be}\sqrt{\hat{\xi}_c} - \be \sqrt{\xi_c} \right) \sqrt{e^{-Q(X_0)} w(X_0)} = 0 ~,
}{eq:con31}
which always vanishes because both configurations satisfy the same boundary conditions
\eq{
\tilde{\be} \sqrt{\hat{\xi}_c} = \be \sqrt{\xi_c} = \beta_c ~.
}{eq:con32}
Therefore, all CDV solutions with given $X_0$ and $\la$ have the same free energy. It follows that the regular CDV solution is only marginally stable against decay into a CDV with conical defect, and vice-versa.

\subsection{Contributions to the Partition Function}
\label{sec:PartitionFunction}

The dominant contributions to the semiclassical partition function from spaces with a conical singularity are given by an integral like \eqref{eq:con102b}. But for systems with a finite cavity the measure in this integral should be treated more carefully. In the canonical ensemble the partition function is expressed as a sum or integral over different internal energies of the system, weighted by the density of states $\exp(S)$ and the Boltzmann factor $\exp(-\beta\,E)$. The internal energy for an ensemble with finite $X_c$ is given by \eqref{InternalEnergy}, which suggests that the appropriate measure is proportional to
\begin{gather}
	\dd E_{c}(\hat{M}) = \frac{\dd \hat{M}}{\sqrt{\hat{\xi}_{c}}} 
\end{gather}
rather than $\dd \hat{M}$. Thus, the semiclassical approximation for the partition function, including contributions from spaces with a conical singularity\,\footnote{This does not include the contributions from smooth fluctuations of the fields around the ground state of the system, which are equally important.}, is
\begin{gather}\label{CanonZ}
	\ZZ \sim \int\limits_{0}^{M^\ts{max}} \dd \hat{M}\,\frac{1}{\sqrt{\hat{\xi}_{c}}} \, \exp\big( - \Gamma_{c}(\hat{M})\big) ~.
\end{gather}
The additional factor of $(\hat{\xi}_{c})^{-1/2}$ in the measure proves to be relevant when computing sub-leading corrections to thermodynamical quantities like the entropy. The functional form of the integrand in \eqref{CanonZ} almost always prevents the direct evaluation of this integral in closed form, but standard semiclassical approximation techniques prove to be very accurate when compared with numerical results.

\subsubsection{The $X_c \to \infty$ limit}

For some theories the conical ensemble exists even as the system is decoupled from the external thermal reservoir. Taking the $X_{c} \to \infty$ limit, or `removing the cavity wall', usually implies $w_c \to \infty$, and in this case the integral \eqref{CanonZ} simplifies.

Provided the limit $X_c \to \infty$ commutes with the integral over $\hat{M}$, the contributions to the semiclassical partition function becomes
\eq{
\ZZ_{\infty} := \lim_{X_c\to\infty} \ZZ \approx \int\limits_0^\infty \dd\hat{M}\,\lim_{X_c\to\infty} \frac{1}{\sqrt{w_c\,e^{Q_c}}} \exp\Big(\,2\pi\,X_{h}(\hat{M}) - \frac{\beta_{c}}{\sqrt{w_c\,e^{Q_c}}}\,\hat{M}\,\Big) ~,
}{eq:con104}
Assuming there are no obstructions, we can use $w(\hat{X}_{h}) = 2\hat{M}$ to convert this into an integral over $\hat{X}_h$
\eq{
\ZZ_{\infty} \approx \frac12 \int\limits_{\hat{X}_{0}}^\infty \dd\hat{X}_h \,  w^\prime(\hat{X}_h)\,  \exp\left(2 \pi \hat{X}_{h} - \frac{1}{2}\, w(\hat{X}_h)\, \lim_{X_c\to\infty} \frac{\beta_{c}}{\sqrt{w_c\,e^{Q_c}}}\right) ~,
}{eq:con106}
where the lower bound is set by the condition $w(\hat{X}_{0}) = 0$, and we have absorbed the factor $(w_c\,e^{Q_c})^{-1/2}$ into the normalization of $\mathcal{Z}_{\infty}$\,\footnote{This may seem odd, since we \emph{just} introduced the factor of $\hat{\xi}_{c}^{-1/2}$ in the measure in \eqref{CanonZ}. The point is that this factor is important for finite $X_c$, but it becomes state-independent, and hence irrelevant, in the $X_c \to \infty$ limit.}. 

Before going further it is important to ask: if $X_c \to \infty$, what is being fixed in defining the ensemble? If $w_c\,e^{Q_c}$ is finite and non-zero in this limit then we may continue to express the action as a function of the same $\beta_c$ that is held fixed at finite $X_c$. But if $w_c\,e^{Q_c}$ diverges then we must take $\beta_c \to \infty$ while keeping the ratio $\beta_c/\sqrt{w_c\,e^{Q_c}}$ finite. In either case, the boundary conditions of the ensemble are specified by fixing a finite, non-zero value for the quantity
\begin{gather}\label{BetaInfty}
	\beta_{\infty} := \lim_{X_c \to \infty} \frac{\beta_{c}}{\sqrt{w_c \, e^{Q_c}}}~.
\end{gather}
In other words, when $X_c \to \infty$ the ensemble is defined by fixing the value of the period, rather than the proper local period at the cavity wall. The contributions to the partition function from conical singularities are then given by
\begin{gather}\label{Zinfty}
	\ZZ_{\infty} \simeq \frac{1}{2}\,\int\limits_{\hat{X}_{0}}^\infty \dd\hat{X}_h \, w^\prime(\hat{X}_h)\,  \exp\left(2 \pi \hat{X}_{h} - \frac{1}{2}\,\beta_{\infty}\, w(\hat{X}_h)\right)~,
\end{gather}
which can also be expressed in the familiar form
\begin{gather}
	\ZZ_{\infty} \simeq \int\limits_{0}^{\infty} \dd\hat{M}\,  \exp\left(S(\hat{M})\right)\,\exp\left(- \beta_{\infty}\, \hat{M}\right) ~.
\end{gather}
Of course, the ensemble only exists if this integral is defined, which requires that $w(X)$ grows sufficiently fast at large values of $X$,
\begin{gather}\label{ExistenceCriteria}
	\lim_{X \to \infty} \frac{w(X)}{X} > \frac{4\pi}{\beta_{\infty}} ~.
\end{gather}
If this condition is satisfied then the ensemble exists. For example, the $X_c \to \infty$ limit is not defined for the Schwarzschild model, which has $w(X) \sim \sqrt{X}$, but it is defined for the Jackiw-Teitelboim model, which has $w(X) \sim X^{2}$. If the large $X$ behavior of $w(X)$ is linear, so that $w(X) = w_{1}\,X + \ldots$ for large $X$, then the ensemble exists only if $\beta_{\infty} > 4\pi/w_{1}$, which corresponds to a Hagedorn temperature $T_{H} = w_{1}/4\pi$. This is especially relevant for the stringy black holes considered in section \ref{subsec:stringy}.

In most cases the integral \eqref{Zinfty} is easier to work with than the integral \eqref{CanonZ}, though approximation methods or numerical techniques are usually still required. The behavior of this integral depends on the ground state of the system, which is determined as in the finite $X_c$ case. The condition for extremizing the action in the $X_c \to \infty$ limit is
\begin{gather}\label{SmoothInfinity}
	\frac{\partial \Gamma_{\infty}(\hat{X}_{h})}{\partial \hat{X}_{h}} = \frac{1}{2}\,\beta_{\infty}\,w'(\hat{X}_{h}) - 2\pi = 0 ~,
\end{gather}
where $\Gamma_{\infty}(\hat{X}_{h}) = \beta_{\infty}\,w(\hat{X}_{h})/2 - 2\pi\hat{X}_{h}$ is the action in the exponent of \eqref{Zinfty}. As before, this is the usual smoothness condition, so the ground state of the system will either be a smooth black hole or the $\hat{M}=0$ HES solution\,\footnote{Unlike the finite $X_c$ case, where the upper limit $M^\ts{max}$ had to be considered when determining the ground state, the upper limit $\hat{M} \to \infty$ is ruled out by the condition \eqref{ExistenceCriteria}.}. A solution $X_{h}$ of \eqref{SmoothInfinity} is a minimum if
\begin{gather}\label{MinInfinity}
	\frac{\partial^{2} \Gamma_{\infty}(\hat{X}_{h})}{\partial \hat{X}_{h}^{\,2}}\bigg|_{X_{h}} = \frac{1}{2}\,\beta_{\infty}\,w''(X_{h}) > 0 ~,
\end{gather}
which is equivalent to the $X_c \to \infty$ limit of the condition \eqref{CDinequality} for positivity of the specific heat.
If a minimum exists, it must be compared to the action of HES, $\Gamma_{\infty}(X_0) = -2\pi X_{0}$, to determine the ground state. Thus, the ground state of the ensemble is a smooth black hole if there is a solution of \eqref{SmoothInfinity} that satisfies \eqref{MinInfinity} and
\begin{gather}
	\Gamma_{\infty}(X_h) - \Gamma_{\infty}(X_0) < 0 \quad \Rightarrow \quad \frac{w(X_h)}{w'(X_h)} < X_{h} - X_{0} ~.
\end{gather}
Otherwise, the ground state is HES. 

When the ground state of the ensemble is a black hole, the integral \eqref{Zinfty} is comparable to contributions to the partition function from smooth Gaussian fluctuations. To see why this is the case, define a new variable $Y = \hat{X}_{h} - X_{h}$ where $X_h$ is the dilaton at the horizon for the black hole ground state. In the semiclassical approximation the main contributions to \eqref{Zinfty} come from configurations close to the ground state, so the integral may be approximated as
\begin{gather}
	\ZZ_{\infty} \simeq \frac{1}{2}\,\exp(-\Gamma_{\infty}(M)) \int\limits_{-\infty}^{\infty} \ns \dd Y\,w'(X_{h})\,\exp\left( - \frac{2\pi^{2}}{C_{\infty}}\,Y^2 \right) ~,
\end{gather} 
where $C_{\infty} = 2\pi w'(X_h)/w''(X_h)$ is the $X_c \to \infty$ limit of the specific heat \eqref{CD}. Evaluating the integral gives
\begin{gather}\label{ZinftyApprox}
	\ZZ_{\infty} \simeq \exp(-\Gamma_{\infty}(M))\,\frac{\sqrt{2\pi C_{\infty}}}{\beta_{\infty}} ~.
\end{gather}
This is comparable to the contributions from \emph{smooth} Gaussian fluctuations around the black hole ground state, because in both cases the coefficient of the quadratic term in the expansion of the action around the minimum is proportional to $C_{\infty}^{-1}$.

If the ground state of the ensemble is HES, then there are two possible approximations for the integral \eqref{Zinfty}. As before, configurations close to the ground state dominate the integral in the semiclassical approximation. But their contributions depend on the behavior of the function $w(\hat{X}_{h})$ in this region. We will assume, for convenience, that $X_{0} = w^{-1}(0) = 0$. Then if the ground state is HES, the conditions $\Gamma_{\infty}(0) = 0$ and $\Gamma_{\infty}(\hat{M}) > 0$ imply that $w(\hat{X}_{h})  \sim \hat{X}_{h}^{\,\gamma}$ near $\hat{X}_{h} = 0$, with $0 \leq \gamma \leq 1$. For $\gamma<1$, the $-\beta_{\infty} w(\hat{X}_{h})/2$ term dominates the action in this region, and the integral is well-approximated as
\begin{gather}\label{HESApprox1}
	\ZZ_{\infty} \simeq \int\limits_{0}^{\infty} \ns \dd\hat{M}\,\exp(-\beta_{\infty}\,\hat{M}) = \frac{1}{\beta_{\infty}} ~.
\end{gather}
On the other hand, if $w(\hat{X}_{h})$ is approximately linear ($\gamma=1$) near $\hat{X}_{h}=0$ then both terms in the action are relevant, and \eqref{Zinfty} is approximated by
\begin{gather}\label{HESApprox2}
	\ZZ_{\infty} \simeq \int\limits_{0}^{\infty} \ns \dd\hat{X}_{h} \,\frac{w'(0)}{2} \,\exp\left(-\left(\frac{1}{2}\,\beta_{\infty}\,w'(0) - 2\pi \right)\,\hat{X}_{h}\right) = \frac{w'(0)}{\beta_{\infty}\,w'(0) - 4\pi} ~.
\end{gather}
In both cases the contributions to the partition function are generally much smaller than other corrections (for instance, from radiation) for a HES ground state.

\subsubsection{Ensembles with finite $X_c$}

Though the analysis is a bit more complicated for finite $X_c$, the same approach yields reliable approximations for \eqref{CanonZ}. When the ground state of the ensemble is a black hole, expanding the action around its minimum gives
\begin{gather}
	\Gamma_{c}(\hat{M}) = \Gamma_{c}(M) + \frac{2\pi^{2}}{C_{c}}\,Y^{2} + \mathcal{O}(Y^3) ~,
\end{gather}
where $Y = \hat{X}_{h} - X_{h}$, and $C_{c}$ is the specific heat at finite $X_c$ \eqref{CD}. The semiclassical limit implies that the integral is dominated by configurations near the minimum, and their contributions may be approximated as
\begin{gather}
	\ZZ \simeq \exp(-\Gamma_{c}(M)) \int\limits_{-\infty}^{\infty} \ns \dd Y\,\frac{w'(X_{h})}{2\,\sqrt{\xi_{c}}}\,\exp\left( - \frac{2\pi^{2}}{C_{c}}\,Y^2 \right) ~.
\end{gather} 
This gives essentially the same result as \eqref{ZinftyApprox}
\begin{gather}\label{ZfiniteApprox}
	\ZZ \simeq \exp(-\Gamma_{c}(M))\,\frac{\sqrt{2\pi C_{c}}}{\beta_{c}} ~,
\end{gather}
but expressed in terms of the relevant quantities evaluated at $X_c$. When the ground state is HES, the analysis is very similar to the $X_c \to \infty$ case. Configurations near $\hat{M}=0$ dominate the integral, and depending on the behavior of $w(\hat{X}_{h})$ in this region \eqref{CanonZ} is approximated by either \eqref{HESApprox1} or \eqref{HESApprox2}, with $\beta_{\infty}$ replaced by $\beta_{c}$.

For comparison with other approaches that calculate corrections to free energy and entropy it is useful to represent the results above as
\eq{
	F_c = -T_c \log \ZZ  = T_c\,\Gamma_{c}(M) - T_{c}\,\log\left(T_{c}\,\sqrt{2\pi C_{c}}\right) ~.
}{eq:angelinajolie}
Then the entropy takes the form $S = S^{(0)} + S^{(1)}$, where $S^{(0)} = 2\pi X_{h}$ is the contribution from the leading term in the free energy, and
\begin{align}\label{eq:lalapetz}
	S^{(1)} = &\,\,\log\left(T_{c}\sqrt{2\pi C_{c}}\right) + \frac{1}{2}\,\left(\frac{\partial \log C_{c}}{\partial \log T_{c}}\right)_{X_c} + 1 \\
		\simeq & \,\, \frac{1}{2}\,\log\left(C_{c} T_{c}^2\right) + \dots\,\,~. 
\end{align}
The second line gives the leading behavior of $S^{(1)}$ in our semiclassical calculations, which takes the same form as corrections from thermal fluctuations \cite{lali:stat1}. These results also apply to ensembles with $X_c \to \infty$, after making the appropriate replacements of quantities evaluated at the cavity wall.

\section{Black hole examples}\label{sec:3}

So far the discussion has allowed for arbitrary functions $U(X)$ and $V(X)$ in the action \eqref{Actionc}. In this section we apply our results to specific models, discussing their thermodynamic properties and determining the leading contributions to the partition function from configurations with conical singularities.

\subsection{Schwarzschild}
  
The Schwarzschild models, which belong to the so-called `$ab$-family' of dilaton gravities \cite{Katanaev:1997ni}, are motivated by a spherically symmetric reduction of gravity with asymptotically flat boundary conditions from $d+1 \geq 4$ dimensions down to two dimensions. The functions $w(X)$ and $e^{Q(X)}$ for these models take the form
\begin{gather}\label{Schwarzw}
	w(X) = (d-1)\,\Upsilon^{\frac{1}{d-1}}\,X^{\frac{d-2}{d-1}} \\ \label{SchwarzQ} 
	e^{Q(X)} = \frac{1}{d-1}\,\Upsilon^{-\frac{1}{d-1}}\,X^{ - \frac{d-2}{d-1}} ~.
\end{gather}
The constant $\Upsilon$ is given by
\begin{gather}
	\Upsilon = \frac{A_{d-1}}{8\pi\,G_{d+1}} ~,
\end{gather}
where $G_{d+1}$ is the $d+1$-dimensional Newton's constant and $A_{d-1}$ is the area of the unit sphere $S^{d-1}$. It will be convenient to retain factors of $\Upsilon$, even though they could be absorbed into a rescaling of the coordinates. Since the growth of $w(X)$ is sub-linear for large $X$, the existence condition \eqref{ExistenceCriteria} implies that the $X_c \to \infty$ limit is not possible for the Schwarzschild model. Thus, in order to work in the canonical ensemble one must couple the system to a heat bath at finite $X_c$.

Setting \eqref{Schwarzw} equal to $2\,\hat{M}$ gives the value of the dilaton at the horizon,
which can then be applied to \eqref{eq:con20} to obtain the entropy of a configuration with conical singularity 
\begin{align}\label{SchwarzS}
	S(\hat{M}) = & \,\, 2\pi\,\left(\frac{2\,\hat{M}}{(d-1)\,\Upsilon^{\frac{1}{d-1}}}\right)^{\frac{d-1}{d-2}} ~. 
\end{align}
This result takes a more familiar form if we solve \eqref{XPrimeDef} for the dilaton as a function of the coordinate $r$
\begin{gather}\label{SchwarzDilaton}
	X(r) = \Upsilon\,r^{d-1} = \frac{A_{d-1}}{8\pi\,G_{d+1}}\,r^{d-1}~.
\end{gather}
The expression for the entropy becomes one-quarter of the horizon area in Planck units, even for configurations with a conical singularity
\begin{gather}
	S(\hat{M}) = \frac{A_{d-1}}{4\,G_{d+1}}\,r_{h}(\hat{M})^{d-1} ~.
\end{gather}
The internal energy of these configurations is obtained from \eqref{InternalEnergy}. Expressed in terms of the boundary conditions and mass parameter $\hat{M}$, it is
\begin{gather}\label{SchwarzEc}
	E_{c}(\hat{M}) = (d-1)\,\Upsilon^{\frac{1}{d-1}}\,X_{c}^{\frac{d-2}{d-1}}\,\left(1 - \sqrt{1 - \frac{2\,\hat{M}}{d-1}\,\Upsilon^{-\frac{1}{d-1}}\,X_{c}^{-\frac{d-2}{d-1}}} \, \right) ~.
\end{gather}
For the HES solution $\hat{M} = 0$ -- the `hot flat space' of \cite{Gross:1982cv} -- the internal energy is zero. For non-zero values of the mass parameter the result \eqref{SchwarzEc} can be inverted to give $\hat{M}$ as a function of the internal energy in the region $X \leq X_c$
\begin{gather}
 	\hat{M} = \hat{E}_{c} - \frac{\hat{E}_{c}^{\,2}}{2\,\sqrt{X_c}} ~,
\end{gather} 
which relates the ADM mass to the internal energy $E_{c}(\hat{M})$ and the gravitational binding energy $-\hat{E}_{c}^{\,2}/2\sqrt{X_c}$ in the cavity \cite{Grumiller:2007ju}.

\begin{figure}
  \centering 
	\includegraphics[width=4in]{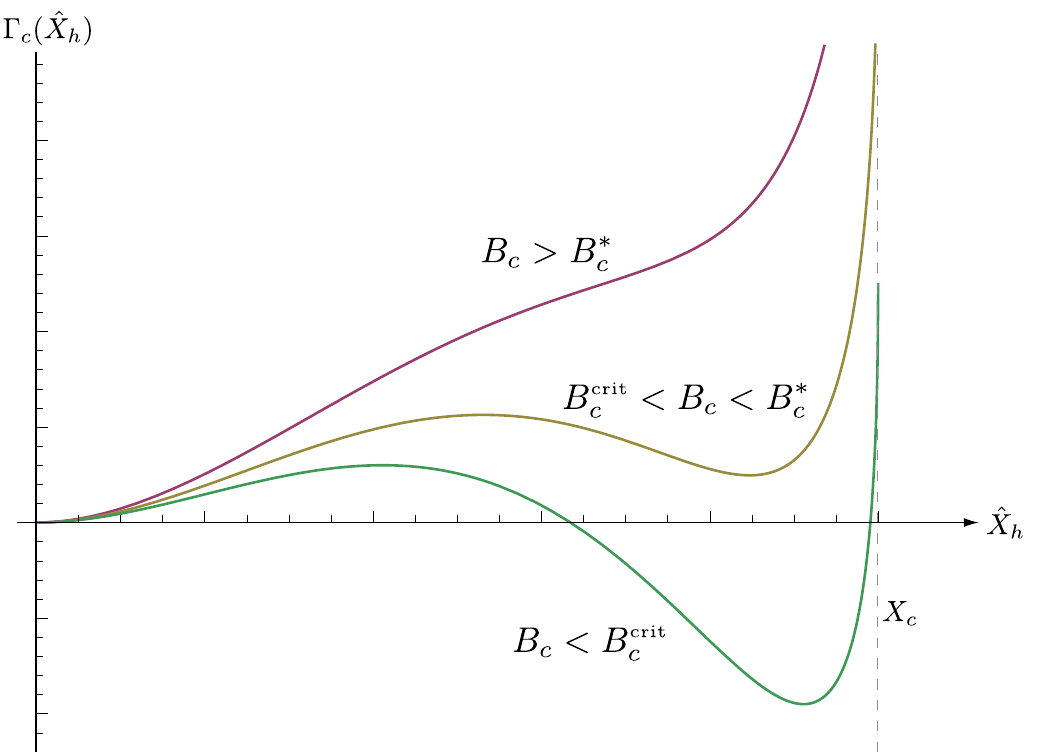}	
  \caption{The Schwarzschild model action $\Gamma_{c}(\hat{X}_{h})$ for $0 \leq \hat{X}_{h} \leq X_c$. Local extrema appear only for $B_c$ below a certain value $B_{c}^{*}$. }\label{fig:Action1} 
\end{figure}

In the rest of this section we consider the phase structure of the Schwarzschild model, and the dominant contributions to the Euclidean partition function coming from black holes with a conical singularity. It is tempting to focus on the familiar example $d+1=4$, but as we will see this is a special case that exhibits qualitatively different behavior than models based on the reduction from $d+1 \geq 5$ dimensions. The analysis is simpler when quantities are expressed as functions of $\hat{X}_{h}$ rather than the mass $\hat{M}$. In that case the action \eqref{eq:gammac} is
\begin{gather}\label{ConActionSchwarz}
	\Gamma_{c}(\hat{X}_{h}) = (d-1)\,\beta_c \,\Upsilon^{\frac{1}{d-1}}\, X_{c}^{\,\frac{d-2}{d-1}} \, \left( 1 - \sqrt{ 1 - \left(\frac{\hat{X}_{h}}{X_c}\right)^{\nts\frac{d-2}{d-1}}} \, \right) - 2\pi\,\hat{X}_{h}~.
\end{gather}
In $d+1=4$ dimensions this is precisely $\beta_c$ times the ``generalized free energy'' obtained by York in \cite{York:1986it}. The action is plotted in figure \ref{fig:Action1} as a function of $\hat{X}_{h}$, for representative values of the boundary conditions. The minimum of the action is either HES or a smooth black hole, depending on the values of $\beta_c$ and $X_c$ fixed by the boundary conditions. The HES solution with $\hat{M}=0$ is always present, but smooth black holes exist in the ensemble only when the smoothness and boundary conditions are both met. This occurs at isolated values of $X_{h}$ that satisfy 
\begin{gather}\label{SchwarzSmoothBC}
		\beta_c = \frac{4\pi\,X_{c}^{\frac{1}{d-1}}}{(d-2)\,\Upsilon^{\frac{1}{d-1}}} \, \left(\frac{X_{h}}{X_{c}}\right)^{\frac{1}{d-1}} \,\sqrt{1 - \left(\frac{X_{h}}{X_{c}}\right)^{\frac{d-2}{d-1}}} ~.
\end{gather}
To analyze this condition it is convenient to define the variables 
\begin{gather}\label{SchwarzVariables}
	\lambda := \left(\frac{X_{h}}{X_{c}}\right)^{\frac{1}{d-1}} \quad \quad \quad B_c := \frac{(d-2)\Upsilon^{\frac{1}{d-1}}}{4\pi}\,\beta_{c}\,X_{c}^{-\frac{1}{d-1}} ~,
\end{gather} 
so that \eqref{SchwarzSmoothBC} takes the form
\begin{gather}\label{SchwarzSmooth}
	B_c = \lambda\,\sqrt{1-\lambda^{d-2}} ~.
\end{gather}
There are no real solutions of \eqref{SchwarzSmooth}, and hence no smooth black holes in the ensemble, for $B_c > B_{c}^{*}$ with  
\begin{gather}
	B_{c}^{*} = \left(\frac{2}{d}\right)^{\frac{1}{d-2}}\sqrt{\frac{d-2}{d}} ~.
\end{gather}
In this case the action \eqref{ConActionSchwarz} is strictly non-negative and the ground state is HES with $\Gamma_c(0) = 0$. But if $B_c < B_{c}^{*}$ there are two black holes in the ensemble, corresponding to two real solutions $0 < \lambda_{-} < \lambda_{+} < 1$ of \eqref{SchwarzSmooth}. The smaller of the two black holes is a local maximum of the action, and the larger black hole is a local minimum. This minimum is positive when $B_c$ is greater than a critical value given by
\begin{gather}
	B_{c}^{\ts{crit}} = \left(\frac{d-2}{d}\right)\,\left(\frac{4\,(d-1)}{d^{2}}\right)^{\frac{1}{d-2}}     ~,
\end{gather}
so that the ground state of the ensemble remains HES for $B_{c}^{\ts{crit}} < B_{c} < B_{c}^{*}$\,. But for boundary conditions that satisfy $0 < B_c < B_{c}^{\ts{crit}}$, the minimum of the action is negative and the ground state of the ensemble is the large black hole.

Using the definition \eqref{SchwarzVariables} and expressing $X_c$ in terms of the radius of the cavity $r_c$, the three regimes of the Schwarzschild model can be described in terms of more conventional variables.
\begin{center}
\begin{tabular}{c|c|c}
Boundary Conditions & \,\, Smooth Black Hole? \,\, & \,\, Ground State\,\,\\
\hline
$T_c < \frac{\sqrt{d(d-2)}}{4\pi\,r_c}\,\left(\frac{d}{2}\right)^{\frac{1}{d-2}}$ & Does not exist & HES \vphantom{\bigg|}\\
$\frac{\sqrt{d(d-2)}}{4\pi\,r_c}\,\left(\frac{d}{2}\right)^{\frac{1}{d-2}} < T_c < \frac{d}{4\pi\,r_c}\,\left(\frac{d^2}{4(d-1)}\right)^{\frac{1}{d-2}}$ & Local minimum & HES \vphantom{\bigg|}\\
$T_c > \frac{d}{4\pi\,r_c}\,\left(\frac{d^2}{4(d-1)}\right)^{\frac{1}{d-2}}$ & Global minimum & SBH \vphantom{\bigg|}\\
\end{tabular}	
\end{center}
The Schwarzschild model has a ``low temperature'' phase, set by the size of the cavity, where smooth black holes do not exist at all -- black holes in the ensemble necessarily exhibit a conical singularity in this regime. Two smooth black holes appear in the ensemble as the temperature is increased at fixed cavity size. One of the black holes is stable against small fluctuations (i.e., $C_{c}>0$), but at intermediate temperatures the system will eventually tunnel from this state to the HES ground state. Finally, a ``high temperature'' phase occurs above a critical temperature that is also set by the size of the cavity
\begin{gather}
	T_{c}^{\ts{crit}} = \frac{d}{4\pi\,r_{c}}\,\left(\frac{d^2}{4(d-1)}\right)^{\frac{1}{d-2}} ~.
\end{gather}
For $T_{c} > T_{c}^{\ts{crit}}$ the ground state of the ensemble is a smooth black hole.

It is worth taking a moment to consider a Gedankenexperiment that examines the phases described above in a ``real world'' setting. Suppose we construct a cavity of macroscopic size in the lab, removing all matter from the interior and holding the walls at a constant temperature. Assuming that gravity is described by the usual Einstein-Hilbert action (and neglecting all physics besides gravity and radiation), what is the relevant ground state for the ensemble? Restoring dimensionful constants, the condition \eqref{SchwarzSmoothBC} becomes
\begin{gather}
	\frac{\hbar\,c}{k_{B}\,T_c} = 4\pi\,r_{h}\,\sqrt{1 - \frac{r_{h}}{r_{c}}} ~.
\end{gather}
The two solutions for a cavity of radius $r_{c} = 0.1\,\text{m}$ held at temperature $T_{c} = 10^{3}\,\text{K}$ are
\begin{gather}
	\frac{r_{h}^\ms{(-)}}{r_c} = 1.8 \times 10^{-6} \quad \quad \quad \frac{r_{h}^\ms{(+)}}{r_c} = 1 - 3 \times 10^{-12} ~.
\end{gather}
The larger solution describes a stable black hole with its horizon about a third of a \emph{picometer} from the wall of the cavity; reasonable laboratory conditions are apparently far into the high-temperature regime! A quick calculation shows that the stable black hole is the ground state of the ensemble, with a free energy of $-10^{48}\,\text{J}$ and an entropy of $10^{68}$. Yet the interior of the cavity remains HES, with a free energy of about $-10^{-4}\,\text{J}$ (from radiation). What prevents the system from tunneling to the overwhelmingly favorable black hole ground state? Recall from the analysis of \cite{Gross:1982cv} and \cite{York:1986it} that the rate of tunneling is approximately $\exp(-\Gamma_{c}(r_{h}^\ms{(-)})/\hbar)$. The action of the unstable black hole is enormous,  $\Gamma_{c}(r_{h}^\ms{(-)}) \sim 10^{56}\,\hbar$\,, so the probability of a tunneling event is, for all intents and purposes, zero\,\footnote{Perhaps a better strategy for studying a black hole of this size is to find one that already exists, build a cavity around it, and then couple the system to a thermal reservoir. However, $r_{h} \simeq 0.1\,\text{m}$ corresponds to a mass well below the Chandrasekhar limit, so the chances of finding one are not good.}.

\begin{figure}
  \centering 
	\includegraphics[width=4.05in]{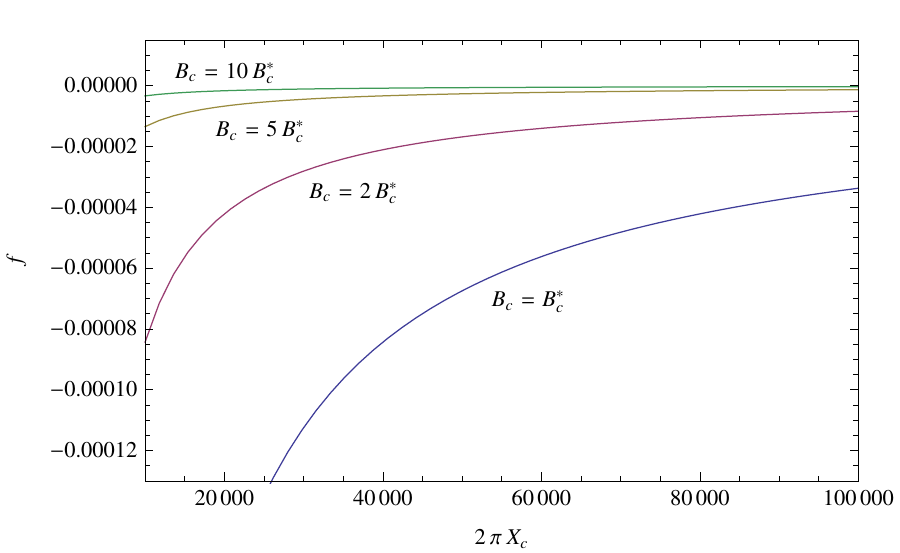}\,\,\,	
	\includegraphics[width=4in]{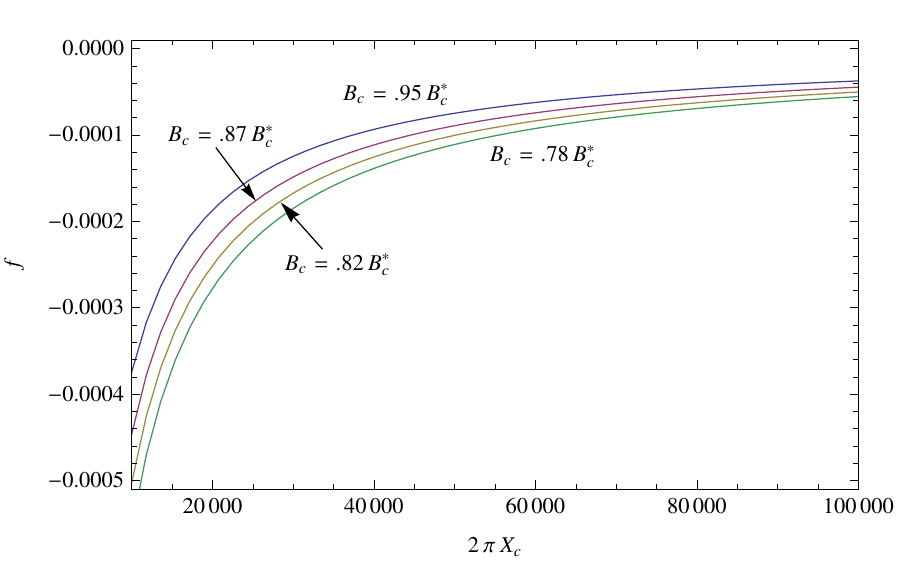}	
  \caption{The fractional error $f = (\ZZ - \ZZ_{\rm num})/\ZZ_{\rm num}$ as a function of $2\pi X_c$, for different values of $B_c$ in the low ($B_c > B_{c}^{*}$) and intermediate ($B_{c}^{\ts{crit}} < B_c < B_{c}^{*}$) temperature regimes.}\label{fig:FELowT} 
\end{figure}

Contributions to the partition function from field configurations with a conical singularity are approximated to high accuracy by relatively simple functions of the boundary conditions, as described in section \ref{sec:PartitionFunction}. For the semiclassical approximation to be valid the action must be very large in units of $\hbar$, which requires $2\pi X_c \gg 1$ in natural units. In the low and intermediate temperature regimes the ground state is HES, and the contributions to the partition function are approximately
\begin{gather}\label{SchwarzCanonZLowT}
	\ZZ(B_c > B_{c}^{\ts{crit}} ) \,\, \simeq \,\, \frac{1}{\beta_c} \, = \,T_c ~. 
\end{gather}
This approximation can be compared with a numerical evaluation of \eqref{CanonZ}. The fractional error, defined as $f = (\ZZ - \ZZ_{\rm num})/\ZZ_{\rm num}$\,, is shown in figure \ref{fig:FELowT} for the case $d+1=4$, with $10^{4} < 2\pi X_c < 10^{5}$ and different values of $B_{c}$. In the low temperature regime the error is typically much less than $10^{-4}$, while in the intermediate temperature regime it is between $10^{-4}$ and $10^{-3}$ for $B_c$ not too close to $B_{c}^\ts{crit}$. The behavior at the critical temperature is described below.

\begin{figure}
  \centering 
	\includegraphics[width=4in]{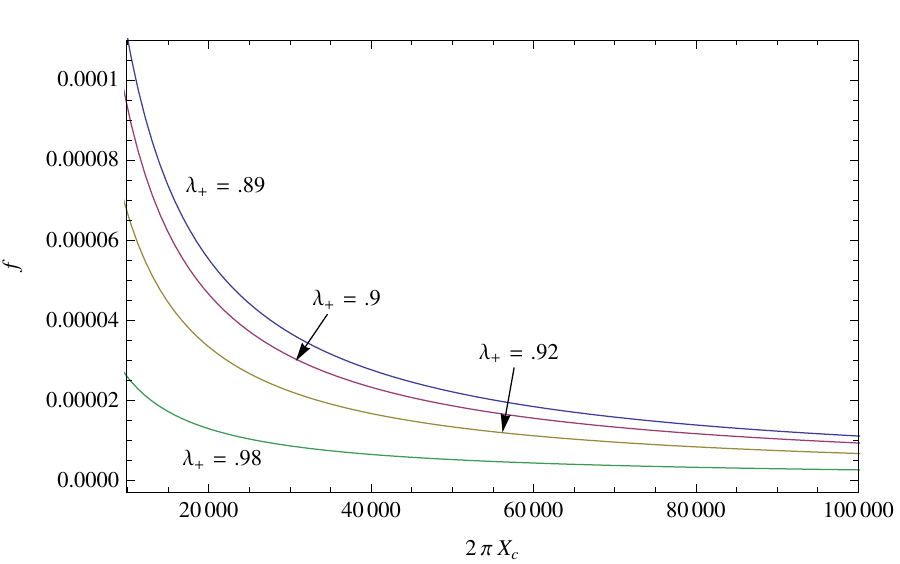} 
  \caption{Fractional error in the approximation for $\ZZ$ as a function of $2\pi X_c$, for different values of $\lambda_{+}$ in the high temperature regime.}\label{fig:FEHighT} 
\end{figure}

In the high temperature phase the behavior of $\ZZ$ is qualitatively different. The approximation \eqref{ZfiniteApprox} for the contributions to the partition function gives 
\begin{gather}\label{SchwarzCanonZHighT}
	\ZZ(B_c < B_{c}^{\ts{crit}}) \,\, \simeq \,\, \exp(-\Gamma_{c}\big(\lambda_{+})\big) \, \frac{(d-2)\sqrt{d-1}\,\lambda_{+}^{\frac{d-3}{2}} X_{c}^{\frac{d-3}{2(d-1)}} }{\sqrt{2\,d\,\lambda_{+}^{d-2} - 4\,}} ~.
\end{gather}
The fractional error for this approximation is shown for the $d+1=4$ model in figure \ref{fig:FEHighT}, with different values of $2\pi X_{c}$ and $\lambda_{+}$. 
In the high temperature regime $\lambda_{+}$ takes values in the range
\begin{gather}\label{SchwarzLambda}
	\left(\frac{4\,(d-1)}{d^2}\right)^{\frac{1}{d-2}} < \lambda_{+} < 1 ~, 
\end{gather}
which becomes $8/9 < \lambda_{+} < 1$ when $d+1=4$. For $2\pi X_c > 10^4$ and $\lambda_{+}>8/9$, the error is typically below about $10^{-4}$. But at $\lambda_{+} = 8/9$ (when $B_c = B_{c}^\ts{crit}$) the error jumps by 1-2 orders of magnitude. This makes sense; at the lower end of \eqref{SchwarzLambda} the smooth black hole has action $\Gamma_{c}=0$, so the ground state of the ensemble is a superposition of the black hole and HES. A better approximation for \eqref{CanonZ} at this transitional value of $\lambda_{+}$ is given by the sum of \eqref{SchwarzCanonZLowT} and \eqref{SchwarzCanonZHighT}. For the $d+1=4$ model this results in a fractional error below $10^{-5}$.

The dominant contribution to the partition function in the high temperature phase is the overall factor of $\exp(-\Gamma_{c}(\lambda_{+}))$. This gives the leading term in the free energy as $F_{c}^{(0)} = \beta_{c}\,\Gamma_{c}(\lambda_{+})$, and the resulting contribution to the entropy for the smooth black hole is
\begin{gather}
	S^{(0)} = 2\pi\,X_{h} = 2\pi\,X_c\,\lambda_{+}^{\,d-1} ~.
\end{gather}
The contributions to $\ZZ$ from configurations with conical singularities give corrections to $F_{c}$ and hence to $S$. The free energy $-T_{c} \log \ZZ$ obtained from \eqref{SchwarzCanonZHighT} is
\begin{gather}
	F_{c} = F_{c}^{(0)} - \frac{(d-2)\Upsilon^{\frac{1}{d-1}}}{4\pi\,X_{c}^{\frac{1}{d-1}}\,\lambda_{+}\sqrt{1-\lambda_{+}^{d-2}}}\,\log\left(\frac{(d-2)\sqrt{d-1}\,\lambda_{+}^{\frac{d-3}{2}} X_{c}^{\frac{d-3}{2(d-1)}} }{\sqrt{2\,d\,\lambda_{+}^{d-2} - 4\,}}\right) 
\end{gather}
which results in an entropy
\begin{align}
	S = & \,\, S^{(0)} + \frac{1}{2} \left(\frac{d-3}{d-1}\right) \log S^{(0)} + \frac{(\lambda_{+}^{d-2}-1)(d\,\lambda_{+}^{d-2} + 2\,(d-3))}{(d\,\lambda_{+}^{d-2} - 2)^{2}} \\ \nonumber
	  & \quad +  \log\left(\frac{d-2}{(2\pi)^{\frac{d-3}{2(d-1)}}}\,\sqrt{\frac{d-1}{2\,d\,\lambda_{+}^{d-2}-4}}\right) ~. 
\end{align}
The last two terms combine to give an $\mathcal{O}(1)$ contribution for all $\lambda_{+}$ in the range \eqref{SchwarzLambda}. Thus, the entropy for the $d+1$-dimensional Schwarzschild model with corrections from conical singularities takes the form
\begin{gather}\label{SchwarzS1}
	S = S^{(0)} + \frac{1}{2} \left(\frac{d-3}{d-1}\right) \log S^{(0)} + \mathcal{O}(1) ~.
\end{gather}
In $d+1=4$ dimensions there is no $\log S^{(0)}$ correction; its absence can be traced back to the dependence of various quantities on the boundary condition $X_c$. In an ensemble that contains smooth black holes, the boundary conditions must satisfy \eqref{SchwarzSmoothBC}, so for fixed $\lambda_{+}$ we have $\beta_{c} \sim X_{c}^{\frac{1}{d-1}}$. The specific heat, on the other hand, scales linearly with $X_c$ at fixed $\lambda_{+}$
\begin{gather}
	C_{c} = \frac{4\pi\,(d-1)\,X_{c}\,\lambda_{+}^{d-1}\,(1- \lambda_{+}^{d-2})}{d\,\lambda_{+}^{d-2} - 2} ~.
\end{gather}
The contributions to $\mathcal{Z}$ involve the factor $\sqrt{C_{c}}/\beta_c$, and for fixed $\lambda_{+}$ this is independent of $X_c$ when $d+1=4$. As a result, the corrections to the free energy and the entropy in that case are $\mathcal{O}(1)$.

It is important to remember that the canonical partition function for the Schwarzschild model is not defined as $X_c \to \infty$, despite the fact that some of the results in this section seem well-behaved in that limit. In the next few sections we will consider models that \emph{do} admit an $X_c \to \infty$ limit. In that case results analogous to \eqref{SchwarzS1} simplify quite a bit, and are easier to interpret.

\subsection{Black holes in AdS}

The spherically symmetric reduction of gravity with a negative cosmological constant gives the AdS-Schwarzschild models. The functions $e^{Q}$ and $w$ in this case are
\begin{align}
	w(X) = &\,\, (d-1)\,\Upsilon^{\frac{1}{d-1}}\,X^{\frac{d-2}{d-1}} + \frac{(d-1)}{\ell^{\,2}}\,\Upsilon^{-\frac{1}{d-1}}\,X^{\frac{d}{d-1}} \\
	e^{Q(X)} = &\,\, \frac{1}{d-1}\,\Upsilon^{-\frac{1}{d-1}}\,X^{-\frac{d-2}{d-1}} ~,
\end{align}
where $\ell$ is the AdS length scale and $\Upsilon= A_{d-1}/8\pi G_{d+1}$. Since $w(X)/X \sim X^{\frac{1}{d-1}}$ at large $X$, this model satisfies the condition \eqref{ExistenceCriteria} for the existence of the partition function in the $X_c \to \infty$ limit. Rather than considering the ensemble with finite $X_c$, we will work with ensembles where the cavity wall is removed\,\footnote{A very nice treatment of the ensemble with finite $X_c$ may be found in \cite{Akbar:2004ke}.}. 

\begin{figure}
  \centering 
	\includegraphics[width=5in]{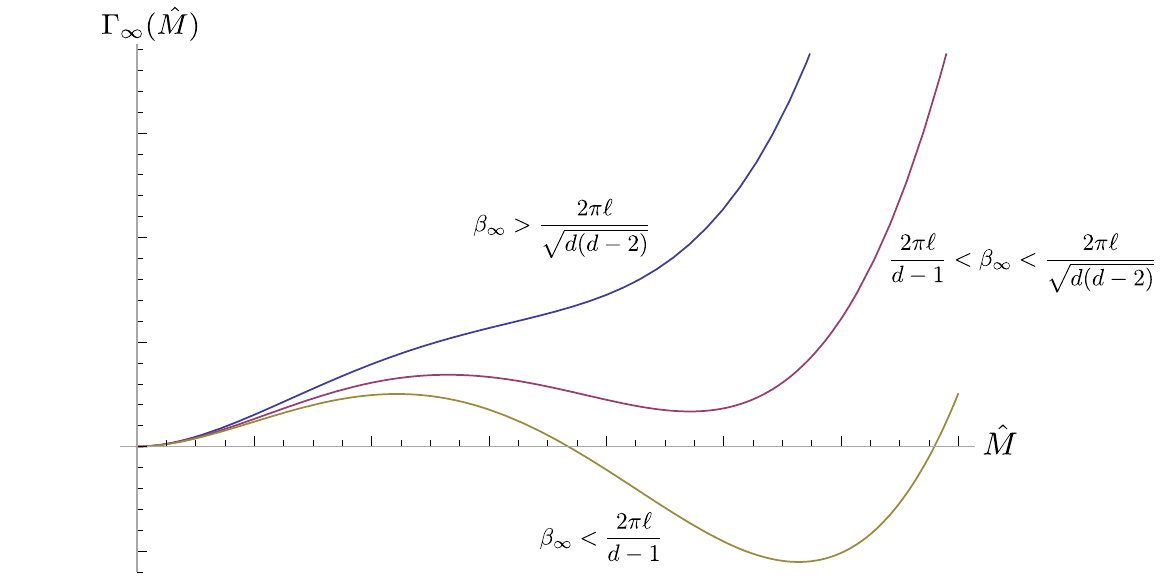}	
  \caption{The action for the AdS-Schwarzschild model with the cavity wall removed.}\label{fig:AdSAction} 
\end{figure}

Following the discussion in section \ref{sec:PartitionFunction}, the ensemble is defined by fixing the period $\beta_{\infty}$. The action for the theory with the cut-off removed is
\begin{gather}\label{AdSAction}
	\Gamma_{\infty}(\hat{M}) = \frac{1}{2}\,\beta_{\infty}\,w(\hat{X}_{h}) - 2\pi \hat{X}_{h} ~.
\end{gather}
Figure \ref{fig:AdSAction} shows plots of the action for representative values of $\beta_{\infty}$. 
The action is extremized by black holes with horizon $X_{h}(M)$ that satisfy the smoothness condition
\begin{gather}
	\beta_{\infty} = \frac{4\pi}{w'(X_{h})} = \frac{4\pi \ell^{\,2}\,\Upsilon^{\frac{1}{d-1}} \, X_{h}^{\frac{1}{d-1}}}{d\,X_{h}^{\frac{2}{d-1}} + (d-2)\,\ell^{\,2}\,\Upsilon^{\frac{2}{d-1}}} ~.
\end{gather}
Expressed as a function of $\beta_{\infty}$, the possible smooth black hole horizons $X_{h}$ are
\begin{gather}\label{AdSHorizon}
	X_{h} = \left(\frac{2\pi \ell^{\,2}}{d\,\beta_{\infty}}\right)^{d-1} \Upsilon \, \left( 1 \pm \sqrt{1 - d(d-2)\,\left(\frac{\beta_{\infty}}{2 \pi \,\ell}\right)^{2}}\right)^{d-1} ~.
\end{gather}
This presents three different scenarios for smooth black holes in the ensemble, assuming $d+1 \geq 4$ (the case $d+1=3$, the BTZ black hole, will be discussed separately). If $\beta_{\infty}>\frac{2\pi\ell}{\sqrt{d(d-2)}}$ then there are no real solutions of \eqref{AdSHorizon}, and all black holes in the ensemble have a conical singularity at the horizon. But if $\beta_{\infty}<\frac{2\pi\ell}{\sqrt{d(d-2)}}$ then there are two smooth black holes, with horizons $0<X_{h}^{-} < X_{h}^{+}$, which correspond to a local maximum ($X_{h}^{-}$, with $C_{\infty}<0$) and minimum ($X_{h}^{+}$, with $C_{\infty}>0$) of the action. The action at the smooth local minimum $X_{h}^{+}$ is
\begin{gather}\label{SmoothAdSAction}
	\Gamma_{\infty}(M) = \frac{2\pi X_{h}^{+}}{d\,(X_{h}^{+})^{\frac{2}{d-1}} + (d-2)\,\ell^{\,2}\,\Upsilon^{\frac{2}{d-1}}}\,\left(\ell^{\,2}\,\Upsilon^{\frac{2}{d-1}} - (X_{h}^{+})^{\frac{2}{d-1}} \right) ~.
\end{gather}
The ground state of the ensemble is easily determined by comparing this to the action $\Gamma_{\infty}(0) = 0$ for the HES solution (the reduction of `thermal AdS'). When $X_{h}^{+} < \ell^{\,d-1}\,\Upsilon$ the action \eqref{SmoothAdSAction} is positive and HES is the ground state of the ensemble; when $X_{h}^{+} > \ell^{\,d-1}\,\Upsilon$ the action is negative and ground state is the smooth black hole. The transition between these phases occurs at a critical value of the period, $\beta_{\infty}^\ts{crit} = \frac{2\pi\ell}{d-1}$, which corresponds to a Hawking temperature $T_{\infty} = \frac{d-1}{2\pi\ell}$. Thus, there are two phases in the model, which can be divided into three distinct temperature regimes. At low temperatures, $T_{\infty} < \frac{\sqrt{d(d-2)}}{2\pi\ell}$, there are no smooth black holes in the ensemble and the ground state is HES. For an intermediate range of temperatures, $\frac{\sqrt{d(d-2)}}{2\pi\ell} < T_{\infty} < \frac{d-1} {2\pi\ell}$, two smooth black holes appear in the ensemble but the ground state remains HES. And finally, at high temperatures, $T_{\infty} > \frac{d-1}{2\pi\ell}$, the ensemble is dominated by the larger of the two smooth black hole solutions.

The approximations derived in section \ref{sec:PartitionFunction} for the contributions to the partition function should be accurate as long as $\ell\,\Upsilon^{\frac{1}{d-1}} \sim \ell/\ell_{pl} \gg 1$. In the low temperature regime where HES dominates the ensemble, $\beta_{\infty} > \frac{2\pi\ell}{\sqrt{d(d-2)}}$\,, the integral \eqref{Zinfty} is approximately
\begin{gather}
	\ZZ_{\infty} \simeq \ell\,T_{\infty}~,
\end{gather} 
independent of the dimension $d+1$ of the original model. The fractional error associated with this approximation is very small (of order $10^{-4}$ or less for $\ell/\ell_{pl} \sim 10^{3}$) and decreases for lower temperatures and larger values of $\ell/\ell_{pl}$. In the phase of the theory dominated by the smooth black hole, $0 < \beta_{\infty} < \frac{2\pi\ell}{d-1}$, the integral \eqref{Zinfty} behaves as
\begin{gather}\label{SBHContributionsAdS}
	\ZZ_{\infty} \simeq \exp\left(-\Gamma_{\infty}(M)\right)\,\left(\frac{\ell}{\ell_{pl}}\right)^{\frac{d-1}{2}} \, \left(\ell\,T_{\infty}\right)^{\frac{d+1}{2}} ~, 
\end{gather}
where the factor in the exponential is given in \eqref{SmoothAdSAction}. At temperatures less than about twice the critical temperature the fractional error in this approximation can be as large as $5\%$ for $\ell/\ell_{pl} \sim 10^{3}$, but at higher temperatures or larger values of $\ell/\ell_{pl}$ this rapidly drops to a fraction of a percent.  

In the high temperature phase the contributions \eqref{SBHContributionsAdS} to the partition function are comparable to corrections from (smooth) quadratic fluctuations around the ground state. The free energy is $F_{\infty} = - T_{\infty} \, \log \ZZ_{\infty}$, so
\begin{gather}\label{SBHEntropyAdS1}
	S = - \frac{\partial F_{\infty}}{\partial T_{\infty}} = 2\pi\,X_{h} + \frac{d+1}{2}\,\log(\ell\,T_{\infty}) + \ldots ~,
\end{gather}
where `$\ldots$' is a constant. At high temperatures, $T_{\infty} \gg \frac{d-1}{2\pi\ell}$, the relation \eqref{AdSHorizon} between the horizon and the temperature is
\begin{gather}
	X_{h} \propto (\ell\,T_{\infty})^{d-1} ~.
\end{gather}
Thus, at high temperatures, the entropy of the smooth black hole can be expressed as 
\begin{gather}
	S = S^{(0)} + \frac{d+1}{2(d-1)}\,\log S^{(0)} ~,
\label{AdSentropy}
\end{gather}
where $S^{(0)} = 2\pi\,X_{h}$ is the dominant contribution to the entropy coming from the free energy of the smooth black hole.

\subsection{Exact Results for the BTZ Black Hole}

The AdS model with $d+1=3$ is an intriguing example where the contributions to the partition function \eqref{Zinfty} can be computed in terms of elementary functions when $X_c \to \infty$. The condition \eqref{AdSHorizon} has a non-zero solution, the BTZ black hole, with horizon
\begin{gather}\label{BTZHorizon}
	X_{h} = \frac{\pi \ell^{\,2}}{2\,G_{3}\,\beta_{\infty}} ~,
\end{gather}
where we have used $\Upsilon = (4 G_{3})^{-1}$ for $d=2$. Thus, unlike higher dimensional models, a smooth black hole exists for ensembles with any value of $\beta_{\infty}$. The action \eqref{AdSAction} for the BTZ black hole is
\begin{gather}
	\Gamma_{\infty}(M) = \frac{1}{8 G_{3} \beta_{\infty}}\,\left(\beta_{\infty}^{\,2} - (2\pi \ell)^{2}\right) ~,
\end{gather}
so HES is the ground state of the ensemble for $\beta_{\infty}$ greater than the critical value  $\beta_{\infty}^\ts{crit} = 2\pi\ell$, and BTZ is the ground state for $\beta_{\infty} < 2\pi\ell$~.

The action \eqref{AdSAction} for a black hole with conical singularity in this model is
\begin{gather}
	\Gamma_{\infty}(\hat{M}) = \frac{1}{8 G_{3}}\,\beta_{\infty}\,\left(1 + \left(\frac{4 G_{3}}{\ell}\right)^{2}\,\hat{X}_{h}^{\,2}\right) - 2\pi \hat{X}_{h} ~. 
\end{gather}
This appears in the integral \eqref{Zinfty} with the measure $\dd \hat{M}$, which can be rewritten using the condition $w(\hat{X}_{h}) = 2\hat{M}$ to give
\begin{gather}
	\dd\hat{M} = \frac{4 G_{3}}{\ell^{2}}\,\hat{X}_{h}\, \dd \hat{X}_{h} ~.
\end{gather}
The integral for contributions to the partition function then takes a form that can be evaluated directly, without the need for approximations
\begin{gather}
	\ZZ \simeq \frac{4 G_{3}}{\ell^{\,2}}\,\exp\left(-\frac{\beta_{\infty}}{8 G_{3}}\right)\,\int\limits_{0}^{\infty} \ns \dd\hat{X}_{h}\,\hat{X}_{h}\,\exp\left(-\frac{2 G_{3}}{\ell^{\,2}}\,\beta_{\infty}\,\hat{X}_{h}^{\,2} + 2\pi \hat{X}_{h} \right) ~.
\end{gather}
Evaluating the integral gives
\begin{gather}
	\ZZ \simeq \frac{1}{\beta_{\infty}}\,\exp\left(-\frac{\beta_{\infty}}{8 G_{3}}\right) + \exp\big(-\Gamma_{\infty}(M)\big)\,\sqrt{\frac{\pi^{3} \ell^{\,2}}{2 G_{3}\,\beta_{\infty}^{\,3}}}\,\left(1 + \text{Erf}\left(\sqrt{\frac{\pi^{2} \ell^{\,2}}{2 G_{3} \beta_{\infty}}}\,\right)\right) ~.
\end{gather}
In the high temperature regime, $T_{\infty} > 2\pi\ell$, the factor of $\exp(-\Gamma_{\infty})$ dominates and $\ZZ$ is
\begin{gather}
	\ZZ_{\infty} \simeq (\ell\,T_{\infty})^{\,\frac{3}{2}}\,\exp\left(-\Gamma_{\infty}(M)\right) ~.
\end{gather}
The entropy of the BTZ is then the usual, leading term, and a sub-leading logarithmic correction with coefficient $3/2$
\begin{gather}\label{BTZentropy}
	S = \frac{\pi^{2} \ell^{\,2}}{G_{3}}\,T_{\infty} + \frac{3}{2}\,\log\left(\frac{\pi^{2} \ell^{\,2}}{G_{3}}T_{\infty}\right) + \ldots
\end{gather}
As with the higher-dimensional AdS models, this is precisely the sort of correction that is obtained when smooth fluctuations around the ground state are included in the path integral. It also agrees with 1-loop calculations, see \cite{Sen:2012dw} and Refs.~therein.

\subsection{The Jackiw-Teitelboim Model}

Another example that can be treated in great detail is the Jackiw-Teitelboim model \cite{Jackiw:1984,Teitelboim:1984}, defined by the functions
\begin{gather}
	w(X) = X^2 \quad \quad \quad e^{Q(X)} = 1 ~.
\end{gather}
It is convenient to work with the location of the horizon, rather than the mass parameter $\hat{M}$. Then the metric function for a black hole with conical singularity is $\hat{\xi} = X^{2} - \hat{X}_{h}^{\,2}$, and the action for such a configuration is
\begin{gather}\label{JTAction}
	\Gamma_{c}(\hat{X}_{h}) = \beta_{c}\,X_c\,\left(1 - \sqrt{1- \frac{\hat{X}_{h}^{\,2}}{X_{c}^{\,2}}}\right) - 2\pi \hat{X}_{h} ~.
\end{gather}
The smoothness condition that extremizes the action yields
\begin{gather}
	\beta_{c} = \frac{2\pi}{X_{h}}\,\sqrt{X_{c}^{\,2} - X_{h}^{\,2}}~.
\end{gather}
Inverting this expression for $X_{h}$ identifies a single smooth black hole that is present in the ensemble for all values of $\beta_c$ and $X_c$
\begin{gather}
	X_{h} = \frac{2\pi \,X_c}{\sqrt{4 \pi^{2} + \beta_{c}^{\,2}}} ~,
\end{gather}
with action
\begin{gather}\label{JTSmoothAction}
	\Gamma_{c}(X_{h}) = X_c\,\left(\beta_{c} - \sqrt{\beta_{c}^{\,2} + 4\pi^{2}}\right)~.
\end{gather}
This expression is always negative, which means that the smooth black hole dominates HES (with action $\Gamma_{c}(0) = 0$). Unlike the previous examples, the conical ensemble for the Jackiw-Teitelboim model always has a smooth black hole for the ground state.

The function $w(X)$ for the Jackiw-Teitelboim model satisfies the condition \eqref{ExistenceCriteria}, which means that the $X_c \to \infty$ limit of the ensemble exists. Before considering the contributions to the partition function for the ensemble with finite $X_c$, let us examine this simpler case. The action for the ensemble with the cavity wall removed is
\begin{gather}
	\Gamma_{\infty}(\hat{X}_{h}) = \frac{1}{2}\,\beta_{\infty}\,\hat{X}_{h}^{\,2} - 2\pi\,\hat{X}_{h} ~, 
\end{gather}
which is minimized by a smooth black hole with $X_{h} =2\pi/\beta_{\infty}$. The action at this minimum is $\Gamma_{\infty}(X_{h}) = - 2\pi^{2}/\beta_{\infty} = -2\pi^{2} T_{\infty}$, and the contributions \eqref{Zinfty} to the partition function are
\begin{gather}
	\ZZ_{\infty} \simeq \int\limits_{0}^{\infty} \ns \dd\hat{X}_{h}\,\hat{X}_{h}\exp(2\pi^{2} T_{\infty}) \,\exp\left(-\frac{1}{2\,T_{\infty}}\,\hat{X}_{h}^{\,2} + 2\pi \hat{X}_{h} - 2\pi^{2} T_{\infty}\right) ~.
\end{gather}
As with the BTZ black hole, this integral can be evaluated in closed form to give
\begin{gather}\label{ZinftyJT}
	Z_{\infty} = \big(2\pi T_{\infty}\big)^\frac{3}{2}\,\exp(2\pi^{2}T_{\infty})\,\left(\frac{1 + \text{Erf}(\sqrt{2\pi^2 T_{\infty}})}{2}\right) + T_{\infty} ~.
\end{gather}
For the semiclassical approximation to hold, the action of the minimum should be large in units of $\hbar$, which requires $T_{\infty} >> 1$ in natural units. In that case the entropy obtained from \eqref{ZinftyJT} is 
\begin{gather}
	S = 4\pi^{2}\,T_{\infty} + \frac{3}{2}\,\log(4\pi^{2} T_{\infty}) ~.
\end{gather}
The first term in this expression is the leading contribution to the entropy of the smooth black hole, and the second term represents a correction due to the contributions from conical singularities. The correction once again coincides with the result \eqref{PartitionFunction} when smooth quadratic fluctuations around the regular ground state are taken into account \cite{Grumiller:2005vy}.

\begin{figure}
  \centering 
	\includegraphics[width=4.5in]{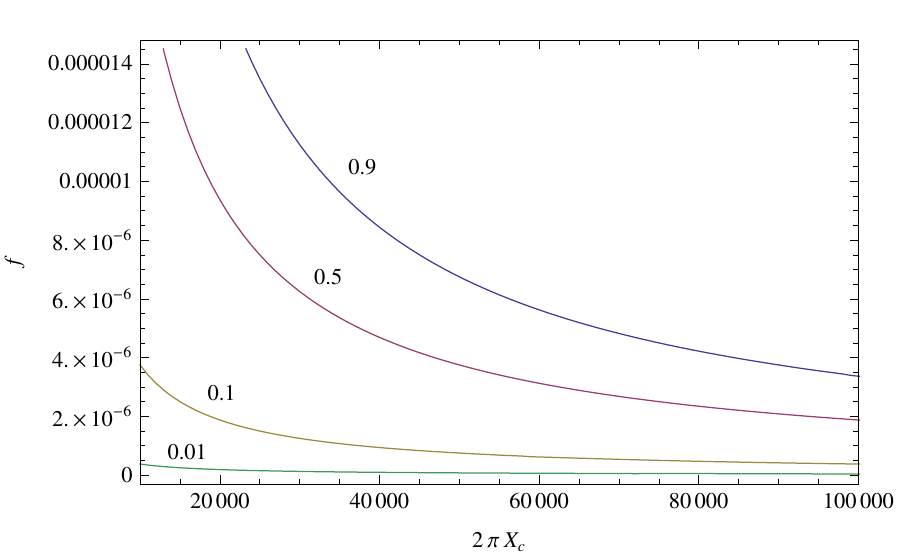}	
  \caption{The fractional error in the approximation for $\ZZ$, as a function of $2\pi X_c$, for different ratios of horizon to cavity size. Each curve is labeled by the value of $X_h/X_c$.} \label{fig:JTFE} 
\end{figure}

For finite $X_c$ the contributions \eqref{CanonZ} to the semiclassical partition function cannot be evaluated exactly, but they are approximated to a high degree of accuracy by a relatively simple function of the boundary conditions. We find
\begin{gather}\label{JTZ1}
	\ZZ_{c} \simeq \exp\big(-\Gamma_{c}(X_{h})\big)\,\frac{1}{(2\pi)^\frac{3}{2} X_c}\,\left(\frac{4\pi^{2}\,X_c\,T_c}{\sqrt{1 + 4\pi^{2} T_{c}^{\,2}}}\right)^{\frac{3}{2}} ~,
\end{gather}
where $\Gamma_{c}(X_{h})$ is the action \eqref{JTSmoothAction} for the smooth black hole. The fractional error for this approximation, compared to a numerical evaluation of \eqref{CanonZ}, is shown in figure \ref{fig:JTFE}.
The entropy calculated using \eqref{JTZ1} consists of a leading term and corrections
\begin{align}\nonumber
	S = &\,\,\left(\frac{\partial}{\partial T_{c}}(T_{c}\,\log\ZZ)\right)_{X_c} \\ \label{JTEntropyXc}
	  = &\,\, \frac{4\pi^{2}X_c\,T_c}{\sqrt{1 + 4\pi^{2} T_{c}^{\,2}}} + \frac{3}{2}\,\log\left(\frac{4\pi^{2}X_c\,T_c}{\sqrt{1 + 4\pi^{2} T_{c}^{\,2}}}\right) + \frac{3}{2\,\left(1 + 4 \pi^{2} T_{c}^{\,2}\right)} +\ldots ~,
\end{align}
where `$\ldots$' indicates terms that are independent of $T_{c}$. The first term in the entropy is the leading behavior associated with the smooth black hole, and the next two terms are the corrections from conical singularity contributions. For small values of $T_c$ \eqref{JTEntropyXc} is approximately
\begin{gather}
	S \simeq \frac{4\pi^{2}X_c\,T_c}{\sqrt{1 + 4\pi^{2} T_{c}^{\,2}}} + \frac{3}{2}\,\log\left(\frac{4\pi^{2}X_c\,T_c}{\sqrt{1 + 4\pi^{2} T_{c}^{\,2}}}\right) 
\end{gather}
which exhibits the same form $S^{(0)} + \frac{3}{2}\log S^{(0)}$ as the ensemble with the cavity wall removed. 
This is not too surprising, since $\beta_{c}/\sqrt{w_c\,e^{Q_c}} = (T_c\,X_c)^{-1}$ must be held fixed in the $X_c \to \infty$ limit, which implies $T_c \to 0$.

\subsection{Stringy black holes}
\label{subsec:stringy}

In this section we consider three models related to black holes that arise in string theory, either as solutions of the $\beta$-functions at lowest order in $\alpha'$, or as exact solutions that incorporate corrections at all orders in $\alpha'$. There is an important difference between this section and the previous ones:  the stringy models must be considered with the cavity wall removed. In string theory one cannot introduce new degrees of freedom in an arbitrary manner (as we have done with the thermal reservoir), and \emph{ad hoc} cut-offs on spacetime fields (the restriction $X \leq X_c$ on the dilaton) are usually equivalent to a truncation of states on the worldsheet that spoil the consistency of the theory.
Since the models we consider all have non-compact target spaces with $X \to \infty$ asymptotically, the $X_c \to \infty$ limit is required for any calculations that are meant to be interpreted in the context of string theory. Additional discussion can be found in \cite{Grumiller:2007ju}. 

It is interesting that there is an intrinsic way to identify string-like behavior within the zoo of two-dimensional dilaton gravity models. Namely, a universal property of all stringy models is that the Weyl-invariant function $w$ is linear in the dilaton field $X$ for large values of the dilaton (which means weak coupling from a target-space perspective). Linearity of $w$ is not just a technical curiosity, but has important physical implications: by virtue of the inequality \eqref{ExistenceCriteria} stringy models always exhibit a Hagedorn temperature. The existence of a Hagedorn temperature (or, equivalently, asymptotic linearity of $w$) therefore can be considered as a defining property of stringy models.

\subsubsection{Witten Black Hole}
 
The Witten Black Hole \cite{Witten:1991yr, Gibbons:1992rh, Nappi:1992as} is obtained from a solution of bosonic string theory with the worldsheet dynamics described by a $SL(2,\mathbb{R})/U(1)$ coset model. When the level of the worldsheet current algebra is taken to be large, the tree-level $\beta$-functions at lowest order in $\alpha'$ have a black hole solution of the form \eqref{metric}. The equations may be obtained from an action with $U(X)$ and $V(X)$ such that
\eq{
w(X)=\la X \qquad e^{Q(X)}=\frac{1}{\lambda\,X} \,,
}{eq:con39}
with the positive parameter $\lambda$ related to the string scale as $\la \sim 1/\sqrt{\alpha'}$. 

The condition \eqref{ExistenceCriteria} implies that the $X_c \to \infty$ limit is only defined for this model if $\beta_{\infty} > 4\pi/\lambda$. In this limit the action is
\begin{gather}
	\Gamma_{\infty}(\hat{X}_h) = 2\pi\,\hat{X}_{h}\,\left(\frac{\lambda\,\beta{_\infty}}{4\pi} - 1 \right)~.
\end{gather}
The smoothness condition that extremizes the action gives a single value of $\beta_{\infty}$ for which a smooth black hole exists:
\begin{gather}
	\beta_{\infty} = \frac{4\pi}{\lambda} ~. 
\end{gather}
This corresponds to the Hagedorn temperature for the model, at which point contributions from states with a conical singularity at the horizon cause the partition function to diverge. Therefore the $X_c \to \infty$ limit of this model does not admit an ensemble containing a smooth black hole. However, one can see from \eqref{eq:con39} that black hole solutions would have scalar curvature of order $1/\alpha'$ near the horizon, indicating that $\alpha'$ corrections are important. In the next section we will consider a model that takes these corrections into account and always has a black hole ground state.

Before moving on it is worth considering two more aspects of the model \eqref{eq:con39}. First, the partition function for this model is well-defined in the $X_c \to \infty$ limit as long as $\beta_{\infty}>4\pi/\lambda$, in which case the ground state is HES. The integral \eqref{Zinfty} can be directly evaluated, and gives
\begin{gather}\label{WittenLowTemp}
	\ZZ_{\infty} = \frac{1}{\beta_{\infty}\,\lambda - 4\pi} ~.
\end{gather}
where $T_{h} = \lambda/4\pi$ is the Hagedorn temperature. Second, despite problems with implementing a finite $X_c$ cut-off in string theory, one might consider this model as an example of a dilaton gravity where the contributions to the partition function can be calculated exactly for finite $X_c$. The action is
\begin{gather}\label{WittenFiniteXcAction}
	\Gamma_{c}(\hat{X}_{h}) = \beta_{c}\,\lambda\,X_{c}\,\left(1 - \sqrt{1- \frac{\hat{X}_{h}}{X_c}}\,\right) - 2\pi \hat{X}_{h} ~,
\end{gather}
with local extrema $X_{h}$ given by the smoothness condition
\begin{gather}
	\beta_{c} = \frac{4\pi}{\lambda}\,\sqrt{1- \frac{X_{h}}{X_c}} ~.
\end{gather}
The ensemble contains a single smooth black hole if $0<\beta_c<4\pi/\lambda$, and the action for this configuration is always negative. Thus, there are two phases: a HES ground state for $\beta_c>4\pi/\lambda$, and a smooth black hole ground state for $0<\beta_c<4\pi/\lambda$. The contributions to the partition function from conical singularities can be calculated exactly in either phase by expressing the action in terms of $\beta_c$ and $\hat{E}_{c}$ 
\begin{gather}
	\Gamma_{c}(\hat{X}_{h}) = \left(\beta_c - \frac{4\pi}{\lambda}\right)\,\hat{E}_{c} + \frac{2\pi}{\lambda^{2}\,X_c}\,\hat{E}_{c}^{\,2} ~.
\end{gather}
Then the integral \eqref{CanonZ} is given in terms of exponentials and error functions. The exact form is not especially enlightening, but the behavior simplifies when $X_c \gg 1$. In the low temperature phase $\beta_c>4\pi/\lambda$ we obtain essentially the same result as \eqref{WittenLowTemp}
\begin{gather}
	\ZZ_{c} \simeq \frac{1}{\beta_{c}\,\lambda - 4\pi} \quad \quad (X_c \gg 1)~.
\end{gather}
At high temperatures, $0< \beta_c < 4\pi/\lambda$, the smooth black hole dominates and the contributions to \eqref{CanonZ} are approximately 
\begin{gather}\label{ZcWitten}
	\ZZ_{c} \simeq \frac{\sqrt{X_c}}{2\sqrt{2}}\,\exp\left(2\pi X_c\,\left(1- \frac{\beta_{c}\,\lambda}{4\pi}\right)^{2}\right)  \quad \quad (X_c \gg 1) ~,
\end{gather}
where the factor in the exponential is minus the action for the smooth black hole. The free energy is $-T_{c}\,\log \ZZ_{c}$, and the resulting entropy is
\begin{gather}\label{FiniteWittenEntropy}
	S = 2\pi X_{h} + \log \left(\frac{\sqrt{X_c}}{2\sqrt{2}}\right) ~.
\end{gather}
Unlike the previous examples, the correction does not appear to be proportional to $\log S^{(0)}$. However, the approximate result \eqref{ZcWitten} assumes $X_c \gg 1$, and this assumption must be treated carefully since the model does not have an $X_c \to \infty$ limit above the Hagedorn temperature $\lambda/4\pi$. The large-$X_c$ result \eqref{FiniteWittenEntropy} can only be trusted if $\beta_c$ remains much less than $4\pi/\lambda$, which implies that $X_h/X_c$ must be close to $1$. Then \eqref{FiniteWittenEntropy} becomes
\begin{gather}\label{FiniteWittenEntropy2}
	S \simeq 2\pi X_{h} + \frac{1}{2}\log (2\pi X_h) + \mathcal{O}(1)~.
\end{gather}
This is, in a sense, the expected result: the functions \eqref{eq:con39} that define this model may also be thought of as the $d\to\infty$ limit of the Schwarzschild model, and \eqref{FiniteWittenEntropy2} is indeed the $d\to\infty$ limit of \eqref{SchwarzS1}. Note, though, that the inverse specific heat for the Schwarzschild model goes to zero in this limit, so perhaps a better interpretation of \eqref{FiniteWittenEntropy2} is the leading term in a $1/d$ expansion for large but finite $d$.

\subsubsection{Exact string black hole}

The black hole background with $\alpha'$ corrections taken into account was studied in \cite{Dijkgraaf:1992ba}. The worldsheet theory is described by an $SL(2,\mathbb{R})/U(1)$ gauged WZW model with level $k>2$. As shown in \cite{Grumiller:2005sq}, the exact string black hole corresponds to a dilaton gravity model with 
\begin{gather}
	w(X) = 2b\,\left(\sqrt{1 + \gamma^{2}} - 1 \right) \quad \quad e^{Q(X)} = \frac{1}{2b\,\left(\sqrt{1 + \gamma^{2}} + 1 \right)} ~,
\end{gather}
where the field $\gamma$ is related to the conventionally defined dilaton $X$ by 
\begin{gather}
	X = \gamma + \sinh^{-1}\gamma ~.
\end{gather}
The parameter $b$ depends on both the level of the model and the string tension 
\begin{gather}
	b = \frac{1}{\sqrt{\alpha'}\sqrt{k-2}} ~.
\end{gather}
For a critical string theory with a target space of dimension $D$, it satisfies the condition
\begin{gather}\label{CentralChargeCondition}
	D - 26 + 6\,\alpha'\,b^{2} = 0 ~.
\end{gather}
Normally this would fix $k$ at a specific value ($k_\ts{crit}=\frac{9}{4}$ for a critical string theory in two dimensions), but as in \cite{Kazakov:2001pj} we will assume that extra matter fields are present that contribute to the central charge. This modifies the condition \eqref{CentralChargeCondition}, which has the effect of allowing us to consider other values of $k$. In practice, $b$ is treated as a fixed parameter and the level takes values in the range $2 < k < \infty$.

The value of the field $\gamma$ at the horizon is related to the level of the CFT by
\begin{gather}
	\gamma_{h} = \sqrt{k(k-2)} \quad \rightarrow \quad X_{h} = \sqrt{k(k-2)} + \sinh^{-1}\big(\sqrt{k(k-2)}\,\big) ~.
\end{gather}
The smoothness condition with the cavity wall removed has solutions for any value of $k>2$
\begin{gather}\label{BetaESBH}
	\beta_{\infty} = \frac{2\pi}{b}\,\sqrt{\frac{k}{k-2}} ~,
\end{gather} 
corresponding to a black hole of mass
\begin{gather}
	M = b\,(k-2) ~.
\end{gather} 
As expected, this black hole is always the ground state of the theory. The on-shell action for the black hole is explicitly negative for all $k>2$
\begin{gather}\label{ESBHaction}
	\Gamma_{\infty}(M) = -\frac{1}{4 G_2}\,\sinh^{-1}\big(\sqrt{k(k-2)}\,\big)~,
\end{gather}
where we have restored factors of the two-dimensional Newton's constant $G_2$. Computing the entropy from the leading term in the free energy $F_{\infty}^{(0)} = T_{\infty}\,\Gamma_{\infty}$ gives
\begin{gather}\label{ESBHentropy0}
	S^{(0)} = \frac{X_{h}}{4 G_2} = \frac{1}{4 G_2}\,\sqrt{k(k-2)} + \frac{1}{4 G_2}\,\sinh^{-1}\big(\sqrt{k(k-2)}\,\big) ~.
\end{gather}
Now, according to the approximation in section \ref{sec:PartitionFunction}, the contributions to the partition function from configurations with a conical singularity give
\begin{gather}
	\ZZ_{\infty} \simeq \exp(-\Gamma_{\infty}(M))\,\frac{\sqrt{2 \pi \, C_{\infty}}}{\beta_{\infty}} ~,
\end{gather}
which suggests a sub-leading correction to the free energy of the form
\begin{gather}
	F_{\infty}^{(1)} \simeq -T_{\infty}\,\log\left(T_{\infty}\,\sqrt{2 \pi\, C_{\infty}}\right) 
	= - T_{\infty}\,\log\left( k^{\frac{1}{4}}\,(k-2)^{\frac{3}{4}}\right) ~.
\end{gather}
and consequently a correction to the entropy of the form
\begin{gather}
	S^{(1)} = \log\left(k^{\frac{1}{4}}\,(k-2)^{\frac{3}{4}}\right) + k - \frac{1}{2} ~.
\end{gather}
Since we are considering a solution of the genus-zero $\beta$-functions, we must take $G_2 \ll 1$ to ensure that the string coupling is small for any value of $k$. Then, in the semiclassical limit $k \gg 1$ the results for $S^{(0)}$ and $S^{(1)}$ simplify
\begin{align}
	S^{(0)} \simeq & \,\, \frac{1}{4\,G_2}\,\big(\,k + \log k\,\big) + \mathcal{O}\left(\frac{1}{G_2}\right)\\
	S^{(1)} \simeq & \,\, k + \log k + \mathcal{O}(1)
\end{align}
We arrive at an interesting result; the corrected entropy takes the form
\begin{gather}
	S \simeq \left(\frac{1}{4\,G_2} + 1\right)\,\left(\vphantom{\Big|} k + \log k  + \mathcal{O}(1)\right) ~.
\end{gather}

\subsubsection{2D type 0A black holes}

The genus-zero beta functions of type 0A string theory have a solution at leading order in $\alpha'$ that describes a two-dimensional black hole with constant Ramond-Ramond flux \cite{Berkovits:2001tg}. This can be viewed as a solution of a model with functions $w$ and $e^{Q}$ given by 
\begin{gather}
	w(X) = \lambda\,X - \lambda\,q^{2}\,\log X \quad \quad e^{Q(X)} = \frac{1}{\lambda\,X} ~,
\end{gather}   
where $\lambda \sim 1/\sqrt{\alpha'}$ is positive, and $q$ is proportional to the flux of each of the two Ramond-Ramond gauge fields\,\footnote{Both Ramond-Ramond gauge fields have the same flux $q_\ts{R}/2\pi\alpha'$, in the conventions of \cite{Douglas:2003up}. This is related to the parameter in $w(X)$ by $q = q_\ts{R}/\sqrt{16\pi}$. }.

The ratio $w(X)/X$ approaches $\lambda$ in the limit $X \to \infty$, indicating that the ensemble is only defined at temperatures below $T_{H} = \lambda/4\pi$ when the cavity wall is removed. This is the same Hagedorn temperature as the Witten black hole model, but unlike that model the ensemble now contains a smooth black hole. The smoothness condition gives
\begin{gather}\label{0ATemp}
	T_{\infty} = T_{H}\,\frac{X_{h} - q^{2}}{X_{h}} ~,
\end{gather}
which identifies a single black hole $X_h$ for any $T_{\infty}$ below the Hagedorn temperature
\begin{gather}\label{0Ahorizon}
	X_{h} = \frac{q^{2}\,T_{H}}{T_{H}- T_{\infty}} ~.
\end{gather}
Combined with the upper limit on the temperature, this result implies that the dilaton at the horizon satisfies $X_h > q^{2}$. Indeed, it turns out that $q^2$ sets a lower bound on the dilaton at the horizon for any configuration in the ensemble, with or without a conical singularity. This is due to the fact that $w(X)$ has a minimum at $X = q^{2}$; the requirement $\hat{\xi}(X) > 0$ for $X > \hat{X}_{h}$ then implies $\hat{X}_{h} \geq q^{2}$.

The action for the smooth black hole is \cite{Davis:2004xi}
\begin{gather}\label{0Aaction}
	\Gamma_{\infty}(X_{h}) = 2\pi\,q^{2}\,\frac{T_{H}}{T_{\infty}}\,\left(1 - \log q^{2} + \log\left(1 - \frac{T_{\infty}}{T_{H}}\right) \right) ~,
\end{gather}
and the action for a configuration with a conical singularity is
\begin{gather}
	\Gamma_{\infty}(\hat{X}_{h}) = 2\pi \hat{X}_{h}\,\frac{T_{H}}{T_{\infty}}\,\left(1 - \frac{T_{\infty}}{T_{H}}\right) - 2\pi\,q^{2}\,\frac{T_{H}}{T_{\infty}} \, \log \hat{X}_{h} ~.
\end{gather}
To determine the ground state, we must compare the action for the smooth black hole to the action for the configuration with $\hat{X}_{h}$ taking the the minimum value $\hat{X}_{h} = q^{2}$. Their difference is
\begin{gather}
	\Gamma_{\infty}(X_h) - \Gamma_{\infty}(\hat{X}_h=q^{2}) = 2\pi q^{2}\,\frac{T_{H}}{T_{\infty}}\,\left(\log\left(1 - \frac{T_{\infty}}{T_{H}}\right) + \frac{T_{\infty}}{T_{H}}\right) ~,
\end{gather}
which is always negative, since the ratio $T_{\infty}/T_{H}$ is less than one. Thus, the ground state of the ensemble is always the smooth black hole \eqref{0Ahorizon}.

Given the action \eqref{0Aaction} for the smooth black hole, the leading contribution to the free energy is $F_{\infty}^{(0)} = T_{\infty}\,\Gamma_{\infty}(X_h)$. The resulting entropy is 
\begin{gather}
	S^{(0)} = 2\pi X_{h} = \frac{2\pi q^{2}\,T_{H}}{T_{H} - T_{\infty}} ~.
\end{gather}
We may now consider corrections from configurations with conical singularities. The integral \eqref{Zinfty} can be evaluated exactly using incomplete gamma functions, but for our purposes the approximation \eqref{ZinftyApprox} is sufficient
\begin{gather}
	\ZZ_{\infty} \simeq \exp(-\Gamma_{\infty}(X_h)) \, \frac{2\pi q\,T_{\infty}^{\,\frac{3}{2}}T_{H}^{\,\frac{1}{2}}}{T_{H} - T_{\infty}} ~.
\end{gather}
The contribution to the free energy is
\begin{gather}
	F_{\infty}^{(1)} = - T_{\infty}\,\log\left(\frac{2\pi q\,T_{\infty}^{\,\frac{3}{2}}T_{H}^{\,\frac{1}{2}}}{T_{H} - T_{\infty}}\right)~,
\end{gather}
and the contribution to the entropy is
\begin{gather}
	S^{(1)} = - \frac{\partial F_{\infty}^{(1)}}{\partial T_{\infty}} = \log\left(\frac{2\pi q\,T_{\infty}^{\,\frac{3}{2}}T_{H}^{\,\frac{1}{2}}}{T_{H} - T_{\infty}}\right) + \frac{T_{\infty}}{T_{H} - T_{\infty}} + \frac{3}{2} ~.
\end{gather} 
As in the previous two examples, the relation between $S^{(1)}$ and $S^{(0)}$ is not clear until we consider the conditions for the semiclassical approximation. With the cavity wall removed, the semiclassical limit is obtained by taking $q^{2} \gg X_{h} \gg 1$. In that case $S^{(1)}$ becomes
\begin{gather}
	S^{(1)} \simeq \frac{1}{2}\,\log S^{(0)} + \mathcal{O}\left(\frac{X_{h}}{q^{2}}\right) ~,
\end{gather}
which is the same general form as the Witten model at finite $X_c$.

\section{Discussion and outlook}
\label{sec:4}

We have considered black holes in an unsuitable box -- a cavity coupled to a thermal reservoir at a temperature that is in general different than the Hawking temperature -- and studied the thermodynamics of a `conical ensemble' that includes these spaces alongside the more conventional smooth field configurations. The focus was on black holes that allow an effective description in terms of 2-dimensional dilaton gravity, including Schwarzschild, Schwarzschild-AdS, Jackiw-Teitelboim and various stringy black holes. We demonstrated that smooth solutions of the equations of motion are locally (perturbatively) stable against singular configurations with a small angular deficit or surplus, proved that the ground state of the conical ensemble never exhibits a conical singularity, and calculated corrections to the entropy and free energy for several pertinent examples.

In all the examples we considered, configurations with a conical singularity result in corrections to the entropy that take the same form as generic logarithmic corrections from thermal (and in some cases also quantum) fluctuations. In fact, our results can be compared with previous results for entropy corrections from these sorts of fluctuations if the following caveats are taken into account:
\begin{enumerate}
	\item In many cases existing results have been obtained in the microcanonical ensemble. Translating our results, derived in the canonical ensemble, into corrections for the microcanonical entropy changes the sign of the coefficient of the logarithmic term.
	\item Matter interactions and non-spherical excitations have been neglected, so naturally we can compare only with results where these contributions are switched off. 
	\item In the Schwarzschild model the partition function does not exist in the $X_c \to \infty$ limit, so our results are only meaningful for a finite cavity. 
\end{enumerate}
With these caveats in mind, we can present the Schwarzschild result \eqref{SchwarzS1} also as a correction to the microcanonical entropy (which coincides at leading order with the canonical entropy $S^{(0)}$)
\eq{
S_{\ts{mc}}^{\ts{Schwarzschild}} = S^{(0)} + \frac{1}{d-1}\,\log S^{(0)}\,\big(C_{\ts{local}} - \tfrac12\,(d-3) + C_{U(1)}\big) ~.
}{eq:Smc}
Here $C_{\ts{local}}$ refers to all matter fields and graviton excitations (basically their contributions to the trace anomaly) and $C_{U(1)}$ is a separate contribution from $U(1)$ gauge fields; our simple approach is not sensitive to either of these contributions. The result \eqref{eq:Smc} agrees precisely with Eq.~(1.4) in \cite{Sen:2012dw} when matter fields are switched off and the graviton excitations are frozen\,\footnote{The ensemble in which that result was calculated is called `mixed ensemble' in the notation of \cite{Sen:2012dw}, but it really corresponds to what we call here `microcanonical', since by construction we neglect angular momentum and are thus only sensitive to the s-wave (or $J=0$) contributions. Therefore, this is the appropriate ensemble to compare with. The 4-dimensional microcanonical result contains an additional contribution coming from the Cartan generators of the rotation group, to which our analysis naturally is blind.}. Likewise, the microcanonical analogs of the entropy corrections \eqref{AdSentropy} for the AdS-Schwarzschild model and \eqref{BTZentropy} for the BTZ black hole are the same as the results obtained in \cite{Das:2001ic}. We consider this to be a consequence of the semiclassical approximation, where the leading corrections to the partition function are given by a Gaussian integral. In both cases -- conical singularities and thermal fluctuations -- the coefficient of the quadratic term in the exponent is proportional to $1/C_c$ (or $1/C_\infty$), leading to similar corrections.

In our analysis we have only taken into account configurations with a single conical defect, located at the black hole horizon. A possible generalization is the inclusion of multiple conical defects, which are not necessarily located at the horizon. This is challenging for at least two reasons. First, the existence and description of multiple conical singularities on a given space is an open problem that depends on curvature bounds and other model-specific quantities \cite{troyanov1991prescribing,Troyanov2007}. Second, the action \eqref{Action} is not suitable for this purpose. We have assumed that field configurations in the ensemble exhibit the same symmetries as the cavity. From the point of view of a higher dimensional theory, the cavity is spherical and elements of the ensemble are spherically symmetric. Including less symmetric configurations in the ensemble requires a more general action that contains terms not present in \eqref{Action}, and ignoring the contributions from these terms leads to nonsensical results. For instance, it is tempting to try and study a conical singularity somewhere between the horizon of a smooth black hole and the cavity wall by replacing $-\hat{X}_{h}\,\alpha$ in \eqref{Actionc} with $-X_{d}\,\alpha$, for some $X_{h} < X_d < X_c$. In any model that admits an $X_c \to \infty$ limit, this gives an action that is unbounded below for $\alpha>0$. But this pathological behavior is simply the result of neglecting important contributions to the action; there is no catastrophic instability that suddenly produces conical singularities at large $X_d$\,.

One obvious shortcoming of our analysis is that it applies only to black holes that are symmetric enough to allow an effective 2-dimensional description in terms of a dilaton gravity model with action \eqref{Action}. It would be of interest to lift our results to higher dimensions, particularly to three and four dimensions. As an example of what one could learn from such a generalization let us focus on three dimensions. A few years ago the interest in 3-dimensional (quantum) gravity was rekindled, see e.g.~\cite{Witten:2007kt,Li:2008dq,Carlip:2008jk,Grumiller:2008qz}. In particular, Maloney and Witten showed that the Euclidean partition function of pure Einstein gravity with a negative cosmological constant is not a sensible CFT partition function and does not factorize holomorphically \cite{Maloney:2007ud}. They arrived at their result by taking into account all known contributions to the Euclidean partition function on the gravity side, assuming smooth metrics, and speculated (among other logical possibilities) that the partition function could be made sensible by taking into account configurations with a conical defect. Given the results of the present work this option does not seem to be likely: we have demonstrated in all explicit examples that the leading contribution from conical defects to the partition function behave in the same way as the leading contributions from thermal or quantum fluctuations. If it remains true in the presence of conical defects that the partition function of 3-dimensional Einstein gravity is 1-loop exact, this means that the qualitative features of the partition function are unlikely to change dramatically upon inclusion of conical defects. It would be of interest to demonstrate this explicitly by lifting our results to three dimensions.

\acknowledgements

We thank Roman Jackiw, Rob Myers and Rafael Sorkin for useful discussions and feedback on an early version of this work. DG is supported by the START project Y435-N16 of the Austrian Science Fund (FWF). RM is supported by a Faculty Research Stipend from Loyola University Chicago. SZ benefits from a PhD research grant of the Institut Interuniversitaire des Sciences Nucl\'eaires (IISN, Belgium); his work is supported by the Belgian Federal Office for Scientific, Technical and Cultural Affairs through the Interuniversity Attraction Pole P6/11. DG and RM thank the Perimeter Institute and the Center for Theoretical Physics at MIT for hospitality and support during the early stages of this work. Finally, RM would like to acknowledge the birth of his wonderful daughter Willa. Her arrival in February 2012 provided him with his best excuse yet for not completing a project on schedule.


\bibliographystyle{apsrev}

\providecommand{\href}[2]{#2}\begingroup\raggedright\endgroup
 

\end{document}